\documentclass[longauth]{aa}  
\usepackage{graphicx}
\usepackage{txfonts}
\begin{document}
   \title{The zCOSMOS Redshift Survey: \\
   the role of environment and stellar mass in
   shaping the rise of the morphology--density relation from z $\sim1$
   \thanks{Based on observations obtained at the European Southern
   Observatory (ESO) Very Large Telescope (VLT), Paranal, Chile, as part of the
   Large Program 175.A-0839 (the zCOSMOS Spectroscopic Redshift Survey)}}

   \author{L. A. M. Tasca\inst{1,2}
 \and J.-P. Kneib \inst{1}
 \and A.~Iovino \inst{3}
 \and O.~Le F\`evre  \inst{1}
 \and K.~Kova\v{c} \inst{4}
 \and M.~Bolzonella \inst{5}
 \and S.~J. Lilly  \inst{4}
 \and R.~G. Abraham  \inst{6}
 \and P.~Cassata \inst{1,7}
 \and O.~Cucciati \inst{1}
 \and L.~Guzzo \inst{3}
 \and L.~Tresse  \inst{1}
 \and G.~Zamorani \inst{5}
  \and P.~Capak \inst{8}
  \and B.~Garilli \inst{2}
  \and M.~Scodeggio \inst{2}
  \and K.~Sheth \inst{9}
  \and D.~Vergani \inst{5}
  \and E.~Zucca \inst{5}
     \and C. M. Carollo \inst{4}
     \and T.~Contini \inst{10}
     \and V.~Mainieri \inst{11}
     \and A.~Renzini \inst{12}
 \and S.~Bardelli \inst{5}
 \and A.~Bongiorno \inst{13}
 \and K.~Caputi \inst{4}
 \and G.~Coppa \inst{5}
 \and S.~de la Torre \inst{1,3,2}
 \and L.~de Ravel  \inst{1}
 \and P.~Franzetti \inst{2}
 \and P.~Kampczyk  \inst{4}
 \and C.~Knobel \inst{4}
 \and A.~Koekemoer \inst{14}
 \and F.~Lamareille \inst{10}
 \and J.~-F.~Le Borgne \inst{10}
 \and V.~Le Brun \inst{1}
 \and C.~Maier \inst{4}
 \and M.~Mignoli \inst{5}
 \and R.~Pello \inst{10}
 \and Y.~Peng \inst{4}
 \and E.~Perez Montero \inst{10,15}
 \and E.~Ricciardelli \inst{16}
 \and J.~D. Silverman \inst{4}
 \and M.~Tanaka \inst{11}
  \and U.~Abbas \inst{1,17}
  \and D.~Bottini \inst{2}
  \and A.~Cappi \inst{5}
  \and A.~Cimatti \inst{18}
  \and O.~Ilbert \inst{1}
  \and A.~Leauthaud \inst{19}
  \and D.~Maccagni \inst{2}
  \and C.~Marinoni \inst{20}
  \and H.~J. McCracken \inst{21}
  \and P.~Memeo \inst{2}
  \and B.~Meneux \inst{13,22}
  \and P.~Oesch \inst{4}
  \and C.~Porciani \inst{23}
  \and L.~Pozzetti \inst{5}
  \and R.~Scaramella \inst{24}
  \and C.~Scarlata \inst{8}
 }
   \offprints{L.A.M. Tasca}

   \institute{Laboratoire d'Astrophysique de Marseille, 
   CNRS-Univerist{\'e} d'Aix-Marseille, 38 rue Frederic Joliot Curie, 
   13388 Marseille Cedex 13, France\\
              \email{lida.tasca@oamp.fr}
         \and
	     INAF-IASF, Via Bassini 15, I-20133, Milano, Italy	 
 \and
 INAF Osservatorio Astronomico di Brera, Via Brera 28, I-20121 Milano, Italy
 \and
 Institute of Astronomy, ETH Zurich, CH-8093, Zurich, Switzerland
 \and
 INAF Osservatorio Astronomico di Bologna, via Ranzani 1, I-40127, Bologna, Italy
 \and
 Department of Astronomy and Astrophysics, University of Toronto, 50 St. George Street, Toronto, ON M5S 3H4, Canada
 \and
 Dept. of Astronomy, University of Massachusetts at Amherst
 \and 
 California Institute of Technology, MC 105-24, 1200 East California Boulevard, Pasadena, CA 91125 USA
 \and
 Spitzer Science Center, 314-6 Caltech, Pasadena, CA 91125, USA
 \and
 Laboratoire d'Astrophysique de Toulouse-Tarbes, Universite de Toulouse, CNRS, 14 avenue Edouard Belin, F-31400 Toulouse, France
 \and
 European Southern Observatory, Karl-Schwarzschild-Strasse 2, Garching, D-85748, Germany
 \and
 INAF - Osservatorio Astronomico di Padova, Padova, Italy
 \and
 Max-Planck-Institut f\"{u}r Extraterrestrische Physik, D-84571 Garching b. Muenchen, Germany
 \and
 Space Telescope Science Institute, 3700 San Martin Drive, Baltimore, MD 21218
 \and
 INAF Osservatorio Astronomico di Torino, Strada Osservatorio 20, I-10025 Pino Torinese, Torino, Italy
 \and
 Dipartimento di Astronomia, Universit´a di Bologna, via Ranzani 1, I-40127, Bologna, Italy
 \and
 Physics Division, MS 50 R5004, Lawrence Berkeley National Laboratory, 1 Cyclotron Rd., Berkeley, CA 94720, USA
 \and
 Centre de Physique Theorique, Marseille, Marseille, France
 \and
 Institut d'Astrophysique de Paris, UMR 7095 CNRS, Universit´e Pierre et Marie Curie, 98 bis Boulevard Arago, F-75014 Paris, France
 \and
 Universit\"ats-Sternwarte, Scheinerstrasse 1, Munich D-81679, Germany
 \and
 Argelander-Institut f\"{u}r Astronomie, Auf dem H\"{u}gel 71, D-53121 Bonn, Germany
 \and
 INAF, Osservatorio di Roma, Monteporzio Catone (RM), Italy
 }

   \date{Received... ; accepted...}

  \abstract
   {For more than two decades we have known that galaxy morphological 
   segregation is present in the Local Universe.
   It is important to see how this relation evolves with cosmic time.
   }
   {To investigate how galaxy assembly took place with cosmic time, 
   we explore the evolution of the morphology--density 
   relation up to redshift z $\sim1$ using about 10000 galaxies drawn from the 
   zCOSMOS Galaxy Redshift Survey. Taking advantage of accurate 
   HST/ACS morphologies from the COSMOS survey, of the well--characterised 
   zCOSMOS 3D environment, and of a large sample of galaxies with spectroscopic 
   redshift, we want to study here the evolution of the 
   morphology--density relation up to z $\sim1$ and its dependence on galaxy 
   luminosity and stellar mass. 
   The multi--wavelength coverage of the field also allows a first study of 
   the galaxy morphological segregation dependence on colour. 
   We further attempt to disentangle between processes that occurred early in
   the history of the Universe or late in the life of galaxies.}
   {The zCOSMOS field benefits of high--resolution imaging in the F814W 
   filter from the Advanced Camera for Survey (ACS). 
   We use standard morphology classifiers, optimised for being
   robust against band--shifting and surface brightness dimming, and  
   a new, objective, and automated method to convert morphological parameters
   into early, spiral, and irregular types. 
   We use about 10000 galaxies down to I$_{AB}=22.5$ with a spectroscopic 
   sampling rate 
   of $33\%$ to characterise the environment of galaxies up to z $\sim1$ from 
   the 100 kpc scales of galaxy groups up to the 100 Mpc scales of the cosmic 
   web. 
   The evolution of the morphology--density relation in different environments 
   is then studied for luminosity and stellar--mass selected, volume--limited 
   samples of galaxies. 
   The trends are described and related to the various physical processes
   that could play a relevant role in the build--up of the morphology--density 
   relation.}
   {We confirm that the morphological segregation is present up to z $\sim1$ 
   for luminosity--selected, volume--limited samples. 
   The behaviour of the morphology--density relation gets flatter 
   at fixed masses expecially above $10^{10.6}$ M$_{\odot}$.
   We suggest the existence of a critical mass above which the physical 
   processes governing galaxy stellar mass also determine the shaping
   of the galaxy more than its environment.
   We finally show that at a fixed morphology there is still a residual
   variation in galaxy colours with density.
   }
   {The observed evolution with redshift of the morphology--density relation 
   offers an opportunity to trace the effect of nature and nurture as
   a function of environment.
   Even though it is based mainly on a biased view,
   the environmental dependence of the morphological evolution
   for luminosity--selected, volume--limited samples seems to indicate that 
   nurture is in play.
   On the other hand, the lack of evolution observed for early--type and spiral
   galaxies that are more massive than $10^{10.8}$ M$_{\odot}$ independents of the 
   environment indicates that nature has imprinted these properties early 
   in the life of these galaxies. 
   We conclude that the relative contribution of nature and nurture in
   different environments strongly depends on the mass of galaxies, 
   consistent with a downsizing scenario.}


   \maketitle
%

\section{Introduction}
\label{sec:intro}

In the standard picture of structure formation, the so--called cold dark matter 
scenario, galaxies form through the cooling and collapse of baryons within dark 
matter haloes (e.g. White \& Rees, 1978).  
Haloes grow in mass through mergers and infall, in a hierarchical fashion: the 
largest objects are formed by mergers of smaller ones.  
Within this general scenario, the details of galaxy formation still must be 
fully understood.  
For example, observations seem to indicate that the most massive galaxies were 
already fully formed at very early times (e.g. Cimatti et al. 2004), 
although numerical models suggest that they in fact have grown until 
today through the 
``dry" accumulation of old populations from accreted galaxies 
(de Lucia \& Blaizot, 2007).  
The fact that these galaxies reside in general within large-scale overdensities 
suggests a connection between the total halo mass and the environment
(Mo \& White, 1996; Lemson \& Kauffmann, 1999).

Strong correlations have been found among the measurable physical properties 
of galaxies:
the visual morphologies of the galaxies well correlate with colour, star
formation rate (SFR), mass, luminosity, surface brightness and the extent to 
which the bulge of a galaxy dominates (Blanton et al. 2003; Roberts \& Haynes, 
1994);
the galaxy stellar mass is closely related to its luminosity;
the surface brightness of giant ellipticals correlates with their size
(Kormendy, 1977);
the luminosity and the rotation velocity of the disk of spiral galaxies are
related by the Tully-Fisher relation (Tully \& Fisher, 1977);
the Fundamental Plane relates luminosity, effective radius and surface 
brightness for elliptical galaxies (Faber \& Jackson, 1976).

Studies to date in the local universe (Kauffmann et al. 1997; Benson et al.
2000, Kauffmann et al. 2004) suggest that observables such as morphology, 
stellar masses, colours and star formation histories in addition to be 
correlated among them are also all correlated to the environment 
(Hogg et al. 2003; Kauffmann et al. 2003, 2004; Blanton et al. 2005a; 
Baldry et al. 2006).
The fraction of galaxies on the red sequence depends strongly on both
stellar mass and environment. In contrast, there are strong indications that 
galaxy stellar masses, independently of the environment, 
determine the colour of a galaxy (Baldry et al. 2006).
The first correlation to be discovered was the one between the environment and 
the galaxy type.
The fact that early--type galaxies are preferentially found in denser regions
than late--type galaxies was first observed by Hubble (1939) and later confirmed
by various studies (Oemler, 1974; Dressler, 1980).

The two major difficulties encountered in early studies were related to the use
of eye--ball morphological classification and projected density estimators.
Using SDSS data Goto et al. (2003) overcome these problems and performed the 
more exhaustive study at low redshift using a
three--dimensional local galaxy density estimation, separating galaxies in four
morphological types and extending the study of the morphology--density relation
into the field regions.
Their results confirmed the existence in the local universe of a morphological 
segregation and support the Postman \& Geller (1984) hypothesis that the
morphology--density relation presents three density ranges of particular
interest or two breaks. 
Postman \& Geller (1984) observed that below a galaxy density of $\sim 5$ 
Mpc$^{-3}$ all population fractions show little density dependence,
while above $\sim 3000$ galaxies Mpc$^{-3}$ the elliptical fraction increases
steeply.
Additionally they connected these regions to physical
mechanisms acting on different time scales upon the galaxy population.

Ram--pressure stripping (Gunn \& Gott, 1972), tidal stripping (Gallagher \&
Ostriker, 1978; Richstone 1976), galaxy mergers and interactions (Toomre, 1977;
White, 1976), galaxy harassment and galactic cannibalism (Ostriker \& Tremaine,
1975; Ostriker \& Hausman, 1977; Hausman \& Ostriker, 1978; Tremaine, 1981) are
some of the environment--dependent physical processes which have been proposed 
as possible drivers of the morphology--density relation.
Some of them (i.e. stripping mechanisms) just act on high density peaks and none 
of them can by itself explain the observed trends.
Initial conditions on galaxy formation could therefore play an important role
too.
Hence the interpretation of the morphology--density relation is not
straightforward and it is still under debate whether the destiny of a galaxy is
decided once its mass is assembled or whether external players related to the
environment have a role, or both.

To perform similar studies at higher redshifts is more challenging due
to the difficulty in estimating the two fundamental quantities
needed: galaxy morphology and overdensity.  
A large sample of faint galaxies with spectroscopic redshift is quite 
difficult to obtain at high redshift, but is necessary to properly estimate 
the environment. 
Intermediate depth surveys with HST combined with ground based
spectroscopy have shown that the Hubble sequence is in place at z $\sim1$ (e.g.,
Lilly at al. 1995; Abraham et al. 1996; Brinchmann et al. 1998; Abraham et al.
2007).
Likewise, deep observations at high resolution were difficult to achieve before
the advent of the Hubble Space Telescope (HST) Advanced Camera for Survey (ACS).
Another fundamental ingredient is a multi--wavelength coverage of the observed
field to provide a uniform photometry to estimate galaxy spectrophotometric
properties, including masses.
Dressler et al. (1997) first studied the morphology--density relation for 10
clusters at z $\sim0.5$ and found that the strength of the relation depends on the
cluster concentration. 
On the specific case of Cl0024+16 at z $=0.4$, Treu et al. (2003) examined the 
environmental
influences on the properties of cluster members from the inner core to well 
beyond the virial radius and found that the fraction of early--type galaxies 
declines steeply from the cluster centre to 1 Mpc radius and more gradually
thereafter, asymptoting toward the field value at the periphery. 
They suggest that mechanisms such as starvation or harassment, operating over
timescales of several Gyr, could explain the mild trends in the morphological 
mix.
Further studies at high--z are mainly based on groups or clusters
(Nuijten et al. 2005; Postman et al. 2005; Poggianti et al. 2008) and the first
relevant attempts to extend this analysis to field galaxies (Smith et al. 2005; 
Capak et al. 2007; Guzzo et al. 2007; Holden et al. 2007 ; 
van der Wel et al. 2007) all make use
of local projected surface densities to estimate the environment.
In parallel, the redshift and luminosity evolution of the 
galaxy colour--density relation up to z $\sim1.5$ has been carefully investigated
in two fundamental works (Cooper et al. 2006; Cucciati et al. 2006).

One problem in comparing the results of these studies at different redshifts 
is the wide range of selection methods and density estimators adopted to 
measure stellar masses and especially morphologies.  
Van der Wel et al. (2007) is the only study where an effort of 
homogenisation has been done by quantifying galaxy morphologies in the local
universe and at higher redshifts in an internally consistent manner.

It is widely accepted that the environment galaxies inhabit should play a 
significant role in their formation and evolution as well as in their star 
formation history.
A detailed understanding of the dependence of the morphological segregation
as a function of mass and luminosity up to z $\sim1$, is therefore an 
important step forward in the challenging astrophysical problem of 
understanding the formation and evolution of galaxies.
The unprecedented extension of the high--resolution HST/ACS imaging coverage in 
the COSMOS field (Scoville et al. 2007; Koekemor et al. 2007) allow us to 
estimate morphologies in a robust and homogeneous way, 
while at the same time sampling a variety of environments over a significant 
redshift range.   
Early analyses of this unique data set were based on the use of accurate 
photometric redshifts to reconstruct the density field at intermediate and 
high densities (Guzzo et al. 2007; Capak et al. 2007), and using 
non--parametric techniques to estimate galaxy morphological types 
(Cassata et al. 2007, Abraham et al. 2007 -- see discussion in the following 
sections).   
In this paper photometric redshifts are only used for the estimation of the
environment combined with the unique information provided by spectroscopic 
information from the first 10k redshifts of the zCOSMOS survey.
This new technique allows us to push our environmental measurements well into 
the low--density regime.  
At the same time, we introduce a revised morphological classification 
extending over the whole redshift range of interest.   
As a result, we can study the morphology--density relation from cluster 
regions to sparse environments and at different redshifts, avoiding internal 
inconsistencies that can be caused by the use of multiple data sets.  

The present work is organised as follows: we introduce the data and the way the
luminosity and mass--selected, volume--limited subsamples are created in Section 2;
a brief 
description of the environment, luminosity and stellar mass estimates as well as
of the galaxy morphology computation is presented in Section 3. 
The implications of our results, shown in Section 4 and 5, for the general
picture of galaxy formation and evolution are discussed in Section 6. 
Throughout this paper, unless otherwise stated, we assume a concordance 
cosmology with $\Omega_M = 0.25$, $\Omega_{\Lambda} = 0.75$ and 
$H_0= 70$ kms$^{-1}$Mpc$^{-1}$. All magnitudes are quoted in the AB system.
   
%

\section{The data}
\label{sec:data}


\subsection{zCOSMOS overview} 
\label{sec:subd1}

zCOSMOS (Lilly et al. 2007) is a large spectroscopic survey undertaken in the 
COSMOS (Scoville et al. 2007a) field using 600 hr of observation with the VIMOS 
spectrograph (Le F\`evre et al. 2003) on the 8 m UT3 "Melipal" of the European
Southern Observatory's Very Large Telescope (ESO-VLT).
VIMOS is a multi--slit imaging spectrograph that can simultaneously observe four
quadrants of roughly $7 \times 8$ arcmin$^2$ each, separated by a cross--shaped
region 2 arcmin wide. 
Slit masks were prepared for the four quadrants of VIMOS using
the VMMPS software (Bottini et al. 2005).
The zCOSMOS redshift survey consists of two parts. 
The brighter, lower redshift component, the so--called zCOSMOS "bright", has a 
pure magnitude selection at $I_{AB} < 22.5$, similar to previous surveys (e.g.
Lilly et al. 1995; Le F\`evre et al. 1995, 2005) and it covers the whole 
1.7 deg$^2$ COSMOS field observed with the Hubble Space Telescope (HST) using 
the Advanced Camera for Surveys (ACS).
The selection criteria for the zCOSMOS--bright survey is based on F814W 
total magnitudes derived fron the 0.1 arcsec resolution HST images
(Koekemoer et al. 2007).
After applying a pure magnitude selection in the range $15 \le I_{AB} \le 22.5$,
we are left with a parent catalogue of $\sim40,000$ objects, the so--called 40k 
sample. 
The 40k sample, even if primarly built from HST/ACS imaging, has an additional 
cleaning done by comparison with $i^{\ast}$ images obtained 
with MEGACAM on the 3.6 m Canada--France--Hawaii telescope. 
Some objects lost in ACS images because of masking of bright stars and their
spikes were recovered and included with their CFHT magnitude since CFHT images 
are rotated by about 10 degrees with respect to the ACS ones.
The higher redshift part, the zCOSMOS "deep", is a survey aiming at observing 
$\sim10,000$ galaxies selected through colour--selection criteria to have 
$1.4 < z < 3.0$, within the central COSMOS 1 deg$^2$.


\subsection{The 10k zCOSMOS redshift sample} 
\label{sec:subd2}

For this work we use the zCOSMOS--bright survey. 
The very impressive deep multi--band photometry in the COSMOS
field (Taniguchi et al. 2007; Capak et al. 2007) is a key component to zCOSMOS
measurements of masses, absolute magnitudes and photometric redshifts, for which
an accuracy of 0.023(1+z) is achieved to $I_{AB} < 22.5$ (Oesch et al. 2009, in 
preparation).
zCOSMOS has currently observed about $50\%$ of the brighter, 
magnitude limited part of the survey. 
This is the so--called 10k bright sample and represents the
data set used in this work. The magnitude selection at $I_{AB} <
22.5$ yields redshifts in the range $0.1 < z <1.2$. To follow the strong
spectral features around $4000 \AA$ to as high redshift as possible, the medium
resolution ($R \sim 600$) grism in the red (5550--9650 \AA) is used. 
The integration time, set to 1h, allows to secure redshifts with a high success 
rate and with a velocity accuracy of $\sim 100$ km s$^{-1}$. 
Roughly 80 masks, distributed over an area of $\sim 1.5$ deg$^2$, have been
already reduced using the VIPGI software (Scodeggio et al. 2005). 
The high reliability of the redshift estimate is assured first by an automatic
determination through the EZ software (Garilli et al. 2009, in preparation) 
and by the reduction of each pointing done independently by scientists in two 
different institutes which participate in the zCOSMOS survey. 
A successive redshift reconciliation is finalised by the assignment of a 
confidence class (see Lilly et al. 2009).
An additional decimal place in the class indicates the consistency between the
spectroscopic and photometric redshifts. It can take the following values: 5, if
the photometric redshift is consistent with the spectroscopic one within
$|z_{phot} - z_{spec}| < 0.008 \times (1+z_{spec})$; 4, if $z_{phot}$ is not
available, usually for stars or quasars; 3, only used for class=9 objects,
indicating that the degeneracy $[OII]\lambda3727-H\alpha$ is solved; and 1, if
the photometric redshift is inconsistent with the spectroscopic one, i.e.,
$|z_{phot} - z_{spec}| > 0.008 \times (1+z_{spec})$.
In this study we consider only galaxies with a secure redshift.
This choice provides a spectroscopic confirmation rate of $99\%$ and
represents approximately $85\%$ of the whole sample.
The mean sectroscopic sampling rate of the 10k bright sample is around
$\sim30\%$, one object out of three in the parent photometric catalogue 
(what we called 40k sample) has a reliable redshift.
The target rate of zCOSMOS objects does not depend on the size, 
brightness or redshift; a large fraction of stars was excluded from the
spectroscopic sample according to their photometry and spectral energy
distribution (SED).

Various selection criteria are applied to select a high quality spectroscopic 
sample to be used in the study of the evolution of the morphology--density
relation. In addition to the flag selection already described, 
broad--line AGNs and residual stars as well as objects with problems in the 
photometry (less than $2\%$) are also removed. 
The requirement of morphological information does not lead to a significant
additional reduction of our sample (less than $\sim 3\%$) .
The final selected sample consists of 8245 galaxies.

%
\begin{table}
\caption{Selection of volume--limited subsamples with evolving--luminosity.}             
\label{table:Tab1}      
\centering                          
\begin{tabular}{l c}        
\hline\hline                 
Luminosity limits &  z intervals\\    
\hline                        
   m1 = -18.5 -z   &	$0.2 < z < 0.5$ \\	  
   m2 = -19 -z     &	$0.2 < z < 0.5$ \\
   m3 = -19.5 -z   &	$0.2 < z < 0.8$ \\
   m4 = -20 -z     &	$0.2 < z < 0.8$ \\
   m5 = -20.5 -z   &	$0.2 < z < 0.8$ \\
   m6 = -21 -z     &	$0.2 < z < 1.0$ \\
\hline                                   
\end{tabular}
\end{table}
%

   \begin{figure}
   \centering
     \resizebox{\hsize}{!}{\includegraphics{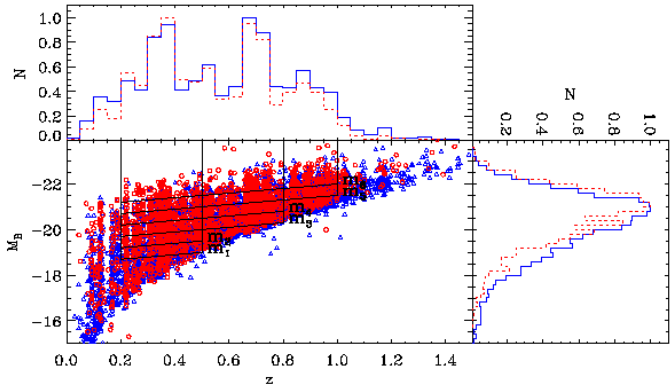}}
       \caption{Central panel: the relation between absolute B--band magnitudes
        and redshifts is shown as a function of the galaxy morphological type
	for the 10k zCOSMOS "bright" galaxy catalogue. 
	Red empty dots represent early--type and blue empty triangles late--type galaxies.
	The black boxes identify the limits for the subsample selection. 
	Top panel: normalised distribution of early (red dotted line) and 
	late--type (blue solid line) galaxies as a function of redshift. 
	Right panel: normalised distribution of early  (red  dotted line) and 
	late--type (blue solid line) galaxies as a function of $M_B$ absolute 
	magnitude. 
       }
      \label{fig:Fig2}
   \end{figure}


\subsection{Evolving--luminosity and stellar mass--selected volume--limited samples} 
\label{sec:subd3}

   \begin{figure}[ht!]
   \resizebox{\hsize}{!}{\includegraphics{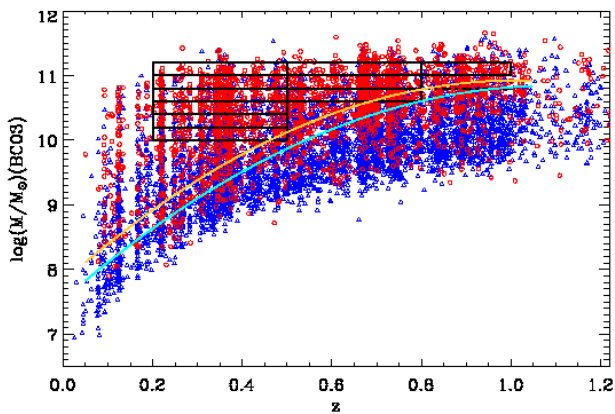}}
      \caption{Redshift distribution of galaxy stellar masses for the 10k 
        zCOSMOS "bright" galaxy catalogue. 
        The relation between stellar mass and redshift is shown for different 
	galaxy morphological types. 
        Red empty dots represent early-type and blue empty triangles late-type 
	galaxies. The black boxes identify the limits for the subsamples 
	selection (see text for more details). 
	The two curves represent the adopted fit of the mass limits computed
	for the Bruzual \& Charlot (2003) models for the early (orange line)
	and late (cyan solid line) galaxy population.
        }
         \label{fig:Fig5}
   \end{figure}

Given the flux limited selection of the zCOSMOS--bright survey we do not 
sample the same absolute magnitude interval at increasing redshifts. 
Moving towards higher redshifts, objects are selected in a progressively 
brighter magnitude range. 
We therefore defined evolving luminosity--selected, volume--limited samples in 
order to study the luminosity dependence of the morphology--density relation 
up to z $\sim1$ being free from incompleteness.  
We allow for one magnitude of passive evolution between z $\sim1$ and present
with the aim to select a homogeneous mix of galaxies at all redshifts, 
consistent with measurements in deep surveys 
(Lilly et al. 1996; Zucca et al. 2006). 
We are aware that the choice of one magnitude for the galaxy luminosity
evolution is an approximate but fair correction. 
A more accurate value would hide our limitation in the estimation of the exact
value, due to our knowledge in galaxy formation.
The relation between galaxies B--band absolute magnitudes and redshifts as well 
as the normalised distribution of early and late--type galaxies as a function 
of absolute B--band magnitude and redshift are shown in Figure~\ref{fig:Fig2}.
The black boxes in the central panel of this figure identify the redshift and 
absolute magnitude bins
chosen in our analysis in such a way that our luminosity volume--limited 
subsamples are not biased against a specific morphological type.
The luminosity limits which assure the completeness in the considered redshift
intervals are summarised in Table~\ref{table:Tab1}. 

Likewise, when studying the mass dependence of the morphology--density relation, 
it is essential to assure that the subsamples used are not biased against a 
specific morphological type.
For this purpose we define the limiting stellar mass as the mass a galaxy
would have if its I--band luminosity is rescaled to the zCOSMOS selection limit 
$I_{lim} = 22.5$, $log(M_{lim})=logM + 0.4 \times (I_{obs} - I_{lim})$
(see Pozzetti et al. 2009, for more details).
As a consequence the distribution of the limiting stellar masses reflects the
distribution of the stellar M/L ratio at each redshift in our sample.
Different morphological types are affected in a different way by incompleteness 
due to the apparent magnitude selection of the survey and the scatter in the 
mass--luminosity relation: late--type galaxies dominate at lower masses in 
each redshift bin, where the completeness of early--type galaxy starts to 
decrease.
Figure~\ref{fig:Fig5} presents the relation between galaxies stellar masses and
redshifts for different morphological types.
The overplotted black boxes represent the mass and redshift bins selection used 
in the analysis.        
Also shown are two curves, corresponding to a $95\%$ completeness level in 
M/L ratio observable at the limit of the survey, which represent the fit of
the limiting mass for the early and late--type population in the redshift 
range $0.1 < z < 1.0$
(i.e. $95\%$ of the objects have a mass limit below this curve).
These conservative selections and our direct tests make us confident that 
the final sample is not affected by selection effects
which would bias our results and conclusions. 

The use in this analysis of luminosity--selected, volume--limited samples is mainly
driven by the aim of studying the luminosity dependence of the 
morphology--density relation and comparing, when possible, our results with 
previous works.
Volume--limited, stellar--mass selected samples have the advantage of being a more 
physically motivated approach. 
Stellar masses are expected to evolve moderately during galaxy life, even if 
galaxies continuously increase in mass through merging or star formation.
This is confirmed by both numerical simulations (de Lucia et al. 2006) and
observational studies (Pozzetti et al. 2007).
We therefore decided to use the accurate stellar mass estimation in the zCOSMOS 
survey to investigate the evolution of the galaxy morphological mix
and its dependence on the environment for volume--limited stellar--mass selected
samples. 
We additionally explore whether the trends which exist for luminosity 
volume--limited samples are still present when using stellar--mass selected, 
volume--limited samples.

\begin{figure*}[hcbt!]
   \centering
   \resizebox{!}{6cm}{\includegraphics{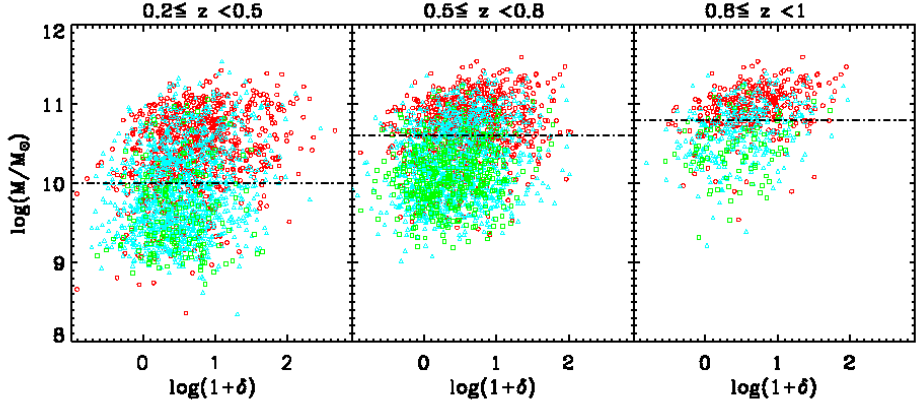}}
      \caption{Mass versus galaxy overdensity as a function of redshift. 
      The logarithm of the stellar masses measured using the BC03 models is 
      shown as a function of the mass--weighted galaxy overdensity computed 
      using the $5^{th}$ nearest neighbour density estimator with a volume 
      limited tracer. 
      The relation is investigated in the three redshift intervals considered in
      the paper.
      Red empty dots represent early-type galaxies, cyan empty triangles spirals 
      and green empty squares irregulars. 
      The black horizontal dotted lines indicate the stellar mass limits 
      adopted in the respective redshift bins.
      It appears that the great majority of irregulars and a relevant fraction
      of spirals are missed when applying a stellar mass selection on top of an
      evolving--luminosity volume--limited selection. 
      }
     \label{fig:Fig4}
   \end{figure*}

It is important to stress, in view of the interpretation of the results shown
later on,  that the use of luminosity--selected, volume--limited samples or
stellar--mass selected, volume--limited samples implicitly
involve the study of different galaxy populations.
This is clearly shown in Figure~\ref{fig:Fig4} where galaxy stellar masses are
plotted as a function of galaxy overdensity in three redshift intervals and
for different morphological types.
The galaxies plotted in the three panels correspond to the objects entering our
luminosity selection, while the horizontal lines represent the mass selection.
The reader should note that low mass late--type objects, in
particular irregular galaxies, do not enter the mass--selected samples, since we
are incomplete at these low masses, while they are included in the luminosity 
selected samples.
A comparative study among results obtained from these two samples allows us 
to identify the morphological class responsible of the observed trends.

\section{Data analysis}
\label{sec:analysis}

We describe in section \ref{sec:suba1} the new method we introduce to
provide an objective and robust morphological classification of all galaxies
considered for this study.
We then briefly review in section  \ref{sec:suba2} how environmental information
are derived from our data
and in section \ref{sec:suba3} how absolute magnitudes and masses are estimated.
We invite the reader to refer to Kova\v{c} et al. (2009), Zucca et al. (2009),
Pozzetti et al. (2009) and Cassata et al. (2009, in preparation) for
a detailed overview of these measurements.


\subsection{The morphological classification} 
\label{sec:suba1}

Galaxy morphology is a key diagnostic of galaxy evolution. As present
cosmological surveys at high redshift such as COSMOS now include more than one 
million of galaxies, it becomes essential to have robust tools for automatic
morphological and structural classifications.
A quantitative scheme should also be objective and reproducible.
Motivated by these needs we developed a classification scheme based on 3
widely adopted nonparametric diagnostics of galaxies structures (Abraham et al.
2003; Lotz et al. 2004), further refining the technique applied in 
Cassata et al. (2007): the concentration index $C$,
the asymmetry parameter $A$ and the Gini coefficient $G$ with the addition of
the galaxy magnitude in the I band.

The concentration is determined using the procedure outlined in Abraham et
al. (1994, 1996a) and it quantifies the central density of the galaxy light
distribution.
It is defined as the logarithmic ratio of the apertures containing all and
$30\%$ of the galaxy light
\begin{equation}
C=5log\frac{r_{all}}{r_{30}}
\end{equation}
The asymmetry parameter quantifies the degree of rotational symmetry of
the galaxy's light distribution.
It is computed by rotating each galaxy by $180^{\circ}$ about its centre, 
subtracting
the rotated image from the original, and dividing the sum of the absolute value
of the pixels in the residual image by the sum of pixel values in the original
image. A correction for background noise is also applied. 
Specifically $A$ is given by
\begin{equation}
A=\frac{\sum_{x,y}|I_{(x,y)}-I_{180(x,y)}|}{2\sum|I_{(x,y)}|}-B_{180}
\end{equation}
where $I$ is the galaxy flux in pixel $(x,y)$, $I_{180}$ is the flux in the
pixel $(x,y)$ rotated by  $180^{\circ}$ and $B_{180}$ is the average asymmetry
of the background. Values span the range 0--1 and larger values correspond to
higher intensity of the residuals;

The Gini coefficient was first introduced in Abraham et al. (2003). It
provides a quantitative measure of the inequality  with which galaxy's light is
distributed among its constituent pixels. To first approximation it can be
seen as a type of concentration index that does not rely on any underlying
symmetry of the galaxy and does not require a well defined galaxy centre.
Values of G span the range from 0 (if the galaxy light is uniformly distributed
among galaxy's pixels) to 1 (if all the light is concentrated in few bright
pixels). To be more precise, after ordering the pixels by increasing flux 
values, G is given by
\begin{equation}
G=\frac{1}{\bar{X}n(n-1)}\sum_{i=1}^{n}(2i-n-1)X_{i}
\end{equation}
where $n$ is the number of pixels of a galaxy and $\bar{X}$ is their mean 
value (Glasser 1962).

To test the consistency of the measured parameters, these quantities have been 
computed using two independent morphological analysis codes adopting a
"quasi--Petrosian" image thresholding technique (Abraham et al. 2007) and a
Petrosian aperture (Lotz et al. 2004, Cassata et al. 2007). 
The quasi--Petrosian isophotes were introduced for the first time in Abraham et 
al. (2007) in order to measure galaxy properties no longer within circular
apertures, whose sizes are multiple of the Petrosian radius (Petrosian 1976), 
but rather on isophotes adapted by an algorithm for galaxies of arbitrary shape.
The quasi--Petrosian definition is therefore more suitable for measuring
structural quantities of galaxies in deep high redshift surveys, considering
the very diverse galaxy populations.

The next step is to convert the galactic structural parameters into morphological
classes. In doing that, we improve the widely--accepted technique of subdividing
the galaxy population into classes on the basis of their position in the
multi--dimensional parameter space.
This technique was first proposed in Abraham et al. (1996) who showed that the 
use of only two parameters, ie. concentration and asymmetry, was sufficient for
separating galaxies among the three main morphological classes (early--type, 
late--type and irregulars). 
Subsequently other authors introduced definitions of
these parameters more robust against surface--brightness selections (Brinchmann
et al. 1998; Wu 1999; Bershaday et al. 2000; Conselice et al. 2000) and
suggested new parameters. The idea of the CAS classification system belongs to
Conselice et al. (2003) who introduced the smoothness ($S$). The Gini
coefficient and the $M20$ moment were proposed recently by Abraham et al. (2003)
and Lotz et al. (2004).
In the standard approach boundaries between classes are generally plotted
manually, errors are not trivial to estimate and the whole procedure lack of
objectivity.
Huertas-Company et al. (2008) recently proposed a generalisation of the 
classical non--parametric classification which uses an unlimited number of
dimensions and non--linear separators, enabling for a simultaneous use of all
the information brought by the different morphological parameters. 
We present here a new algorithm which defines the boundaries in an automated and
objective way and which allows for error estimate.
The main steps of the proposed methodology can be summarised as follows: 

\begin{itemize}

\item we first measure non--parametric morphological and structural parameters 
for all galaxies in the sample;

\item we randomly draw from the main catalogue a sample of $\sim 500$ galaxies 
that we further use as a training set for the automatic classification 
algorithm.
All galaxies in this reference sample are visually classified by at least three
experts (PC, OLF, LT) to reduce misplacement and subjectivity (Naim et al. 2005). 
This training set is then employed to calibrate the volume filled by the
data in the multi--parameter space. This critical step leads to the
choice of the parameter space regions, in which galaxies of a specific
morphological type reside, that are then used for the final classification;

\item we select the morphological parameters which contribute in increasing our
ability to separate the various morphological classes since the information added by a
specific parameter does not always turn out to be an improvement for the
classification.
It should be kept in mind that the simultaneous use of more parameters may 
complicate the interpretation of the properties of galaxies in each class.
We additionally consider non--morphological parameters, such as the luminosity 
or the redshift of the galaxy since morphological parameters might depend on the 
luminosity and redshift of the galaxy (Brinchmann et al. 1998; Bershday et al. 
2000).
We define in this way the parameter space to be used for the automatic
classification where each of the chosen parameters has a specific weight $w$,
computed as the value which optimised the completeness giving the lower possible
contamination;

\item we compute for each galaxy ($i$) in our sample the distance $d_i$ with 
respect to each galaxy in the reference sample according to 
the expression
\begin{equation}
  d_i=\sum \frac{(Q_{ref}-Q_i)^2}{w_Q}  
\end{equation}
where $Q$ it could be equal to $A,C,G,S,$etc...
We stress that in this analysis we only use the $A,C$ and $G$ parameters, 
since they have shown to have the higher capabality to separate morphologies.\\
We then assign a morphological class to the galaxy depending on the most 
frequent class among the ones of the 11th nearest neighbours in the 
multi--parameter space considered;

\item we finally use the algorithm to classify the visually inspected sample. 
The comparison of the visual and automatic classification on the control sample 
allows for error evaluation as well as contamination and completeness 
assessment.

\end{itemize}

   \begin{figure}
   \centering  
   \resizebox{\hsize}{!}{\includegraphics{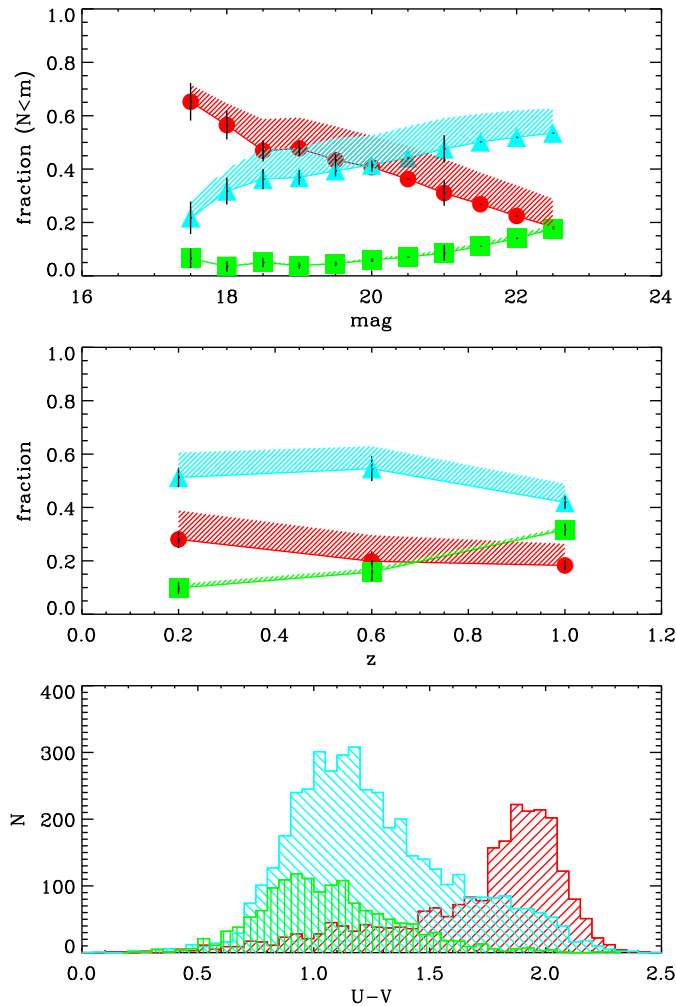}}
     \caption{Top panel: fraction of early (red filled circles), spiral
     (cyan 
     filled triangles) and irregular (green filled squares) types as a function of
     the selection magnitude. 
     Middle panel: fraction of early--type (red filled circles),
     spirals (cyan filled
     triangles) and irregulars (green filled squares) types as a function of
     redshift.
     In the top and middle panels the shaded regions represent the uncertainty 
     in the morphological classification. 
     The vertical error bars are the formal fraction errors given by simple
     binomial statistics. 
     The solid lines are the interpolation obtained using the morphological 
     classification relative to the less contaminated sample.
     Bottom panel: colour distribution of early--type (red $45^{\circ}$--inclined shaded 
     area), spirals (cyan $-45^{\circ}$--inclined shaded area) and irregular galaxies 
     (green vertical shaded area).
     }
      \label{fig:Fig10}
   \end{figure}

The originality of this new galaxy's classification method mainly consists in the
possibility to place galaxies in morphological classes using a multi--parameter
space where the non linear boundaries are defined in an objective and 
reproducible way, which additionally allows for error estimate.
In addition, internal parameters can be optimised to obtain a more complete or
less contaminated sample relative to a specific morphological class.

The described procedure strongly depends on the quality of the imaging.
It is widely recognised that image depth and the resolution are the most 
important ingredient needed to perform a reliable morphological classification.
For the specificity of this study we benefit of the fact that
the COSMOS field was imaged with the Hubble Space Telescope (HST) using the
Advanced Camera for Surveys (ACS). Images were taken through the wide F814W
filter and the reconstructed pixel scale is $0.03"$ pixel$^{-1}$. 
The full width half--maximum of the point--spread function is $0.12"$, yielding 
an unprecedented resolution of small high redshift galaxies. 
The median exposure depth across the field is 2028 seconds.
Leautheaud et al. (2007) show that the completeness of the ACS catalogue is about
$90\%$ for objects with a FWHM of $0.2"$ at F814W=26.6.
Taking into account the flux limited selection of the zCOSMOS survey at I=22.5
we are confident in the completeness of our sample. 

More precisely, for the analysis presented in this paper, we computed 
non--parametric morphological parameters for all objects in the 40k catalogue 
starting from HST/ASC images using two independent codes. 
These codes differ essentially for the type of aperture used: 
"quasi--Petrosian" in one case and Petrosian in the other. 
 A training sample of roughly 500 objects was drawn from the HST/ACS catalogue
fulfilling the luminosity condition $18 < F814W < 24$. 
We performed a detailed eye--ball classification into ellipticals, lenticulars,
spirals of all types (Sa, Sb, Sc, Sd), irregulars, point--like and undefined
sources and we then grouped these classes into early (E,S0), spirals (Sa,Sb,Sc,
Sd) and irregular types. 
It is this coarser classification that is considered in building the
training set.
We then include in the late--type class, galaxies
presenting a spiral or irregular morphology.
The unclassified objects are not used for the training.
We then used this reference set to establish the separation regions in the
multi--parameter space defined by three non--parametric quantities
(concentration, asymmetry, Gini) and the galaxy's apparent magnitude. 
We finally run our algorithm to assign morphological classes to all objects in 
the 10k catalogue with the exception of objects that have entered the catalogue 
but were selected on CFHTLS images.
The main difference we observed in the final classification is in the relative 
fraction of spirals and irregulars provided by the two codes.
Nonetheless since for the purpose of this study spirals and irregulars are
merged together in the late--type class, we are confident that our results are 
independent on the code used.

Figure~\ref{fig:Fig10} shows the general trends for the morphological catalogue
used in this study. 
The reader should observe that the choice of clean or complete samples
would change the fraction of galaxies of a specific morphological type as a 
function of magnitude and redshift, and should additionally be warned that this
fraction is strictly related to our class definition 
(see Cassata et al. 2009, in preparation, for more details).
The top panel shows the fraction of different morphological types as
a function of I--band apparent magnitude.
Three classes are considered: early--types (red filled dots), spirals (cyan 
filled triangles) and irregulars (green filled squares). 
Early--type galaxies are predominant at the brightest magnitudes, where they
represent $\sim 65 \%$ of the whole population. Their contribution rapidly
decreases, reaching $\sim 20 \%$ at the magnitude limit of this study. 
An opposite tendency is observed for the late--type population for which the
fraction
increases towards fainter magnitudes, where spirals are the dominant population. 
In the middle panel we present the evolution of the fraction of the 
morphological classes with redshift. 
We note that the sizeable growth of the fraction of irregulars above $z \sim 0.6$,
balanced by the continue decrease of the elliptical fraction from $\sim30\%$ at 
low redshift to $\sim20\%$ at $z \sim 1$.
The fraction of disk galaxies remains rather constant at $\sim50 \%$, 
with a mild decrease towards higher redshifts. These trends are in agreement
with Conselice et al. (2004) and Cassata et al. (2005).
In the first and second panels error bars are the formal fraction errors given 
by binomial statistics, while the uncertainty in the morphological 
classification is represented by the shaded areas.
The shaded areas are drawn as the regions between measurements computed 
using visually inspected "clean" and "complete" samples. 
For objects in the "clean" sample (filled symbols in Figure~\ref{fig:Fig10}) we
are confident to have low contamination by other morphological types, 
but we know to be somewhat incomplete. 
Instead, the upper edge of the shaded area identifies the limit obtained with a
complete but contaminated sample.
A detailed explanation on how these two samples are created is given in
Section~\ref{sec:specT}. 
Nonetheless we anticipate to the reader that a pre--selection based on the
galaxy SED is applied to decide on galaxies that will be further visually 
inspected.
 
The $U-V$ distribution histograms for early--type galaxies, spirals and
irregulars are show in the bottom panel. 

Morphological k--correction usually affects samples for which the
morphological classification is performed in a single band, that maps
different rest--frame wavelengths at different redshifts. In fact, the
galaxy morphology varies as a function of the observed wavelength,
with objects appearing morphologically later at bluer wavelengths
(Windhorst~et~al.~2002; Papovich~et~al.~2003). However, since the
strongest change in morphology occurs in correspondence of the Balmer
break (i.e. Cassata~et~al.~2005; Sheth et al. 2003, 2008), the strongest bias 
is introduced when optical rest--frame morphologies
($4000 \AA<\lambda<10000 \AA$) are compared with UV ones 
($\lambda<4000 \AA$). 
In our case, the i--band maps the rest--frame r--band at z=0.2, the V--band at 
z=0.6 and the B--band at z=1, therefore galaxies are observed redwards the 
Balmer break confidently up to z$\sim 0.8$. 
Nonetheless, the effects of morphological k--correction are small when just two 
broad classes of early-- and late--type galaxies are used (see
Brinchmann~et~al.~1998), which is the case for a large part of the analysis 
carried out in this paper.
Bluewards the Balmer break red elliptical galaxies, even if faint, still keep 
an early--type morphology.
Instead spirals could be classified as irregulars since bulges are faint 
while bright star formation regions dominate on the disk.

   \begin{figure*}[ht!]
   \centering
   \resizebox{!}{19cm}{\includegraphics{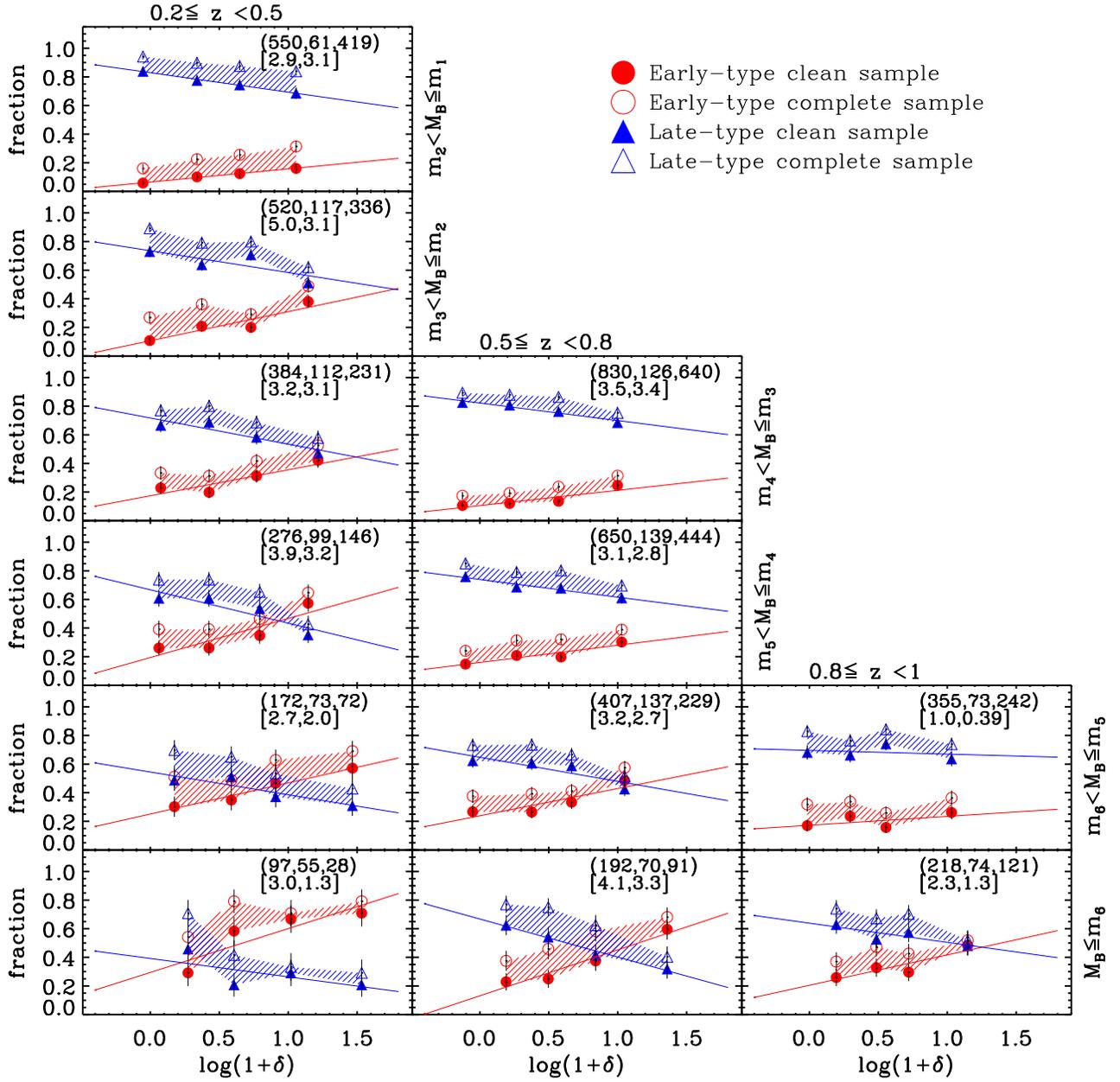}}
     \caption{Luminosity dependence of the morphology--density relation. The panels show the relation 
     between the fraction of galaxies of different morphological types and the mass--weighted 
     galaxy overdensity computed using the $5^{th}$ nearest neighbour density estimator with volume 
     limited tracers. The morphology--density relation is presented in three redshift intervals,
     which increase from the left to the right of the panels (as quoted in the top), 
     and for different absolute magnitude bins (as quoted on Table~\ref{table:Tab1} and reported on the 
     right--end side of the panels).  
     Filled red circles and filled blue triangles represent the more conservative sample of early and late 
     type galaxies, respectively.
     Empty red circles and empty blue triangles represent the more complete, but also contaminated, 
     samples of early and late--type galaxies, respectively.
     All points are plotted as a function of the median overdensity in the respective quartile 
     (see text for more details).
     The shaded areas represent the uncertainty in the morphological classification, while the vertical error
     bars are the standard fraction errors given by binomial statistics.
     Enclosed in parenthesis are the total number of objects in the magnitude and
     redshift bin considered, as well as the number of the more conservative sample of
     early and late--type galaxies.
     Within square brackets we give the significance of the deviation from
     zero of the slope of the linear fits relative
     to the clean early and late--type galaxies respectively. 
     }
      \label{fig:Fig3}
   \end{figure*}
%


\subsection{The density field reconstruction} 
\label{sec:suba2}
We use the density estimates of Kova\v{c} et al. (2009).
The design of the zCOSMOS survey was optimised for providing an accurate
reconstruction of the environments of galaxies on scales ranging from the 100
kpc to the 100 Mpc.
A new algorithm (ZADE, Kova\v{c} et al. 2009) has been developed which 
produces environmental measurements on a broad range of scales with a 
noise reduction in the observed overdensities. 

Cooper et al. (2005) perform numerical experiments indicating that an 
uncertainty smaller than 0.005 in the photometric redshift measurement 
together with the use of projected densities are the minimum requirements
for photometric surveys to properly reconstruct the environment in all 
density regimes. 
However, in their previous investigation of the morphology--density relation 
with the COSMOS data, Guzzo et al. (2007) show that at the intermediate and 
high densities, it is in fact possible to define an unbiased estimator to 
succesfully benefit of the large size of photometric 
surveys.   
The innovative idea of the ZADE method adopted to reconstruct the density 
field used here (Kova\v{c} et al. 2009) is in fact that of combining the 
accuracy of spectroscopic redshift with the large number statistics of 
photometric redshifts.
In particular, the ZADE algorithm modifies the redshift likelihood 
distributions, obtained from photometric redshift codes, using the information 
from the available spectroscopic redshifts of nearby galaxies.
This is done for all the objects in the 40k catalogue not yet spectroscopically 
observed, in order to achieve the number statistics of photometric redshifts.

The environment reconstruction has been performed using various spatial
filters with smoothing kernels of both fixed and adaptive size. 
The nearest neighbour technique as well as the fixed aperture with a top hat
filter or Gaussian weighting methods are implemented.
In particular, two samples of tracers, flux limited and volume--limited, have been
considered to compute simple counts or luminosity and mass--weighted
overdensities.
The overdensities are finally corrected for the edge effect by dividing the 
measured quantity by the fraction of the area used to get the overdensity 
which is within the zCOSMOS survey limits.
We address the reader to Kova\v{c} et al. (2009) for a complete
description.

Among the abundant variety of estimators, we decided to use here the $5^{th}$ 
nearest neighbour approach with volume--limited tracers and mass--weighted 
overdensities. 
This choice allows to explore the smaller scales environments attainable by the
data,
while sampling the same physical scale at all redshifts, since the galaxy
population used as tracer does not change with redshift.
Mass--weighted overdensities are chosen since the mass is expected to be a 
better tracer of the environment than simple galaxy counts, since it is the 
galaxy stellar mass that better traces the density in a given volume.
Since galaxy properties (ie. luminosity, morphology, etc.) are known to be 
related to the galaxy stellar mass, there is the possibility that the use of 
mass-weighted overdensities introduces an additional degree of correlation, in 
the relations with density analysed in this paper, simply because the density 
estimator used is itself dependent on the galaxy stellar mass. 
All the relations shown later on in this paper have been checked using also 
counts overdensities (i.e. no weight), and we are reassured by the fact that 
they are still present, even if sometimes at a slightly lower confidence level. 


\subsection{Stellar masses and absolute magnitudes estimate} 
\label{sec:suba3}

Stellar masses for all galaxies in the current zCOSMOS 10k catalogue are defined
as the integral of the star formation rate, subtracted by the amount of the mass
of gas processed by stars and returned to the ISM during their evolution.
This is estimated by fitting the galaxy spectral energy distribution, as sampled by the
COSMOS multi--$\lambda$ photometry, with a library of stellar population models
based on Bruzual \& Charlot (2003).
The reader is referred to Pozzetti et al. (2007, 2009) for a full description of
the methodology used to derive stellar masses and for a discussion of their
robustness and intrinsic errors.

Absolute magnitudes are computed following the method described in the
Appendix of Ilbert et al. (2005). The K--correction is computed
using a set of templates and all the multi--band photometry data 
available. However, in order to reduce the template dependency, the
rest frame absolute magnitude in each band is derived using the
apparent magnitude from the closest observed band, shifted at the
redshift of the galaxy. With this method, the applied K--correction is
the smallest.
For each galaxy the rest frame magnitudes were matched with the
empirical set of SEDs described in Arnouts et al. (1999),
composed of four observed spectra (CWW, Coleman et al. 1980) and
two starburst SEDs computed with GISSEL (Bruzual \& Charlot 1993).  
The match is performed by minimizing a $\chi^2$ variable on
these templates at the spectroscopic redshift of each galaxy.

Finally, galaxies have been divided into four spectrophotometric types
based on the CWW and starburst templates (more details are given in
Section~\ref{sec:specT}).

%

\section{The morphology--density relation}
\label{sec:mor_den}

At present all the analyses carried out to explore the morphology--density 
relation confirm the Dressler et al. (1980) finding of a morphological 
segregation of galaxies in the local universe. 
This is reflected also in the widely recognised fact that at low redshift 
early-type galaxies are more clustered than later morphological types 
(e.g. Guzzo et al. 1997).

Rather than simply tracing the evolution of the morphology--density relation 
up to z $\sim1$, our goal here is to explore in detail how this evolution 
depends on luminosity and stellar mass.
The environmental trends shown here have been tested against various ways to 
estimate the environment and the result shown is robust and independent of it.


\subsection{Morphology--density relation: luminosity dependence} 
\label{sec:subr1}

   \begin{figure}
   \resizebox{\hsize}{!}{\includegraphics{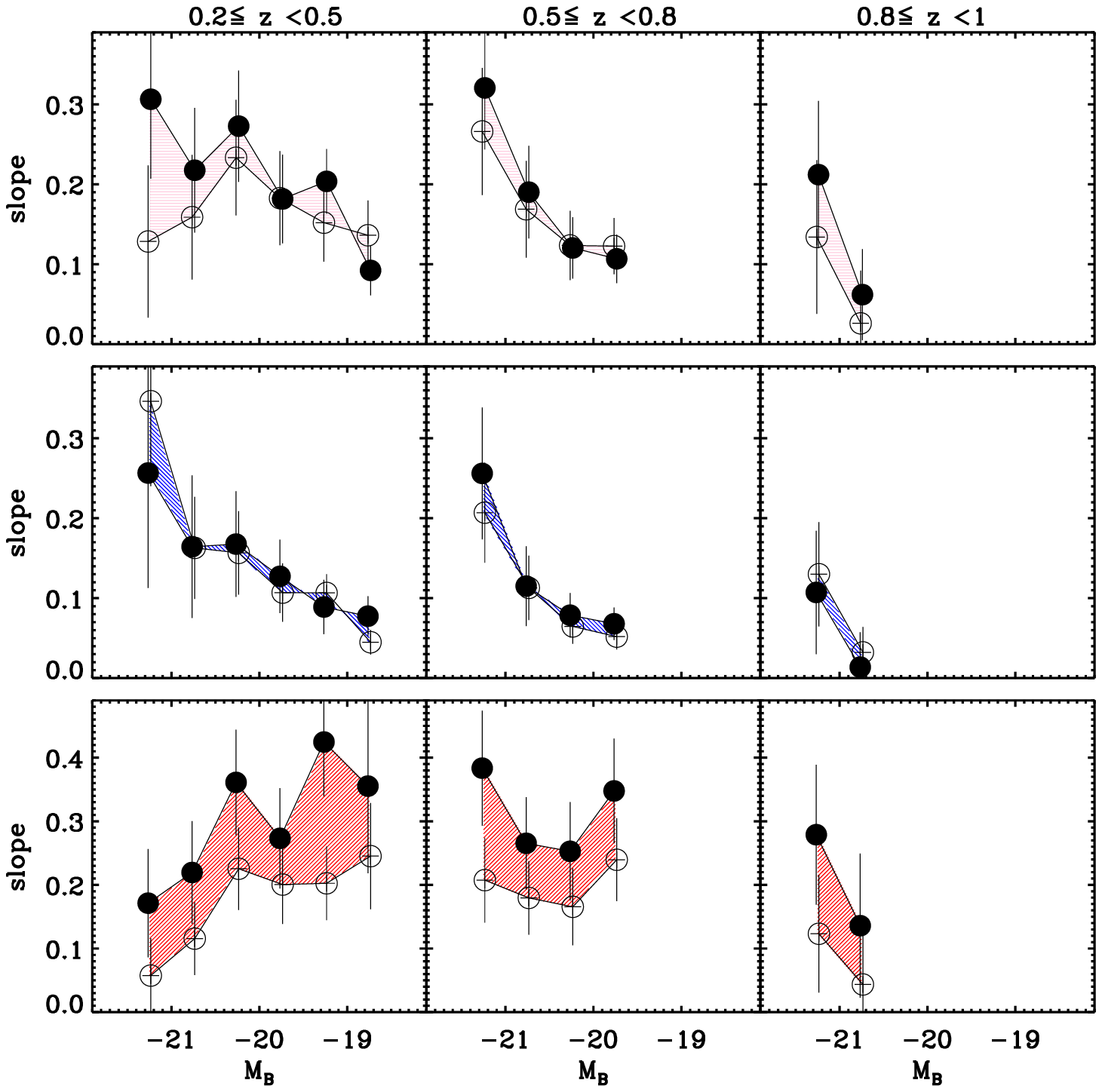}}
     \caption{Slope of the morphology-density relation as a function of 
     galaxy luminosity, in three different redshift intervals ($0.2 \le z <0.5, 
     0.5 \le z < 0.8$ and $0.8 \le z < 1$).
     On the top row each point represents the slope of the linear fit to the 
     morphology-density relation for early--type galaxies already shown in 
     Figure~\ref{fig:Fig3}. 
     The absolute value of the slope for late--type galaxies is the same, 
     since their behaviour with the environment is complementary 
     to the early--type one.
     On the central and bottom row each point represents the slope of the linear
     fit to the morphology--density relation derived when the fraction of 
     galaxies (late--type in the middle row, early--type in the bottom row) is 
     plotted instead in logarithmic scale. 
     The use of a logarithmic scale allows the direct visualisation of 
     fractional changes in the abundance of a population as a function of 
     environment, irrespective of its global abundance.
     In all panels filled circles correspond to the clean samples and empty 
     circles to the complete samples. 
     The shaded areas quantify the uncertainty due to the
     morphological classification.
     }
      \label{fig:Fig15}
   \end{figure}

   \begin{figure*}[ht!]
   \centering
   \resizebox{!}{19cm}{\includegraphics{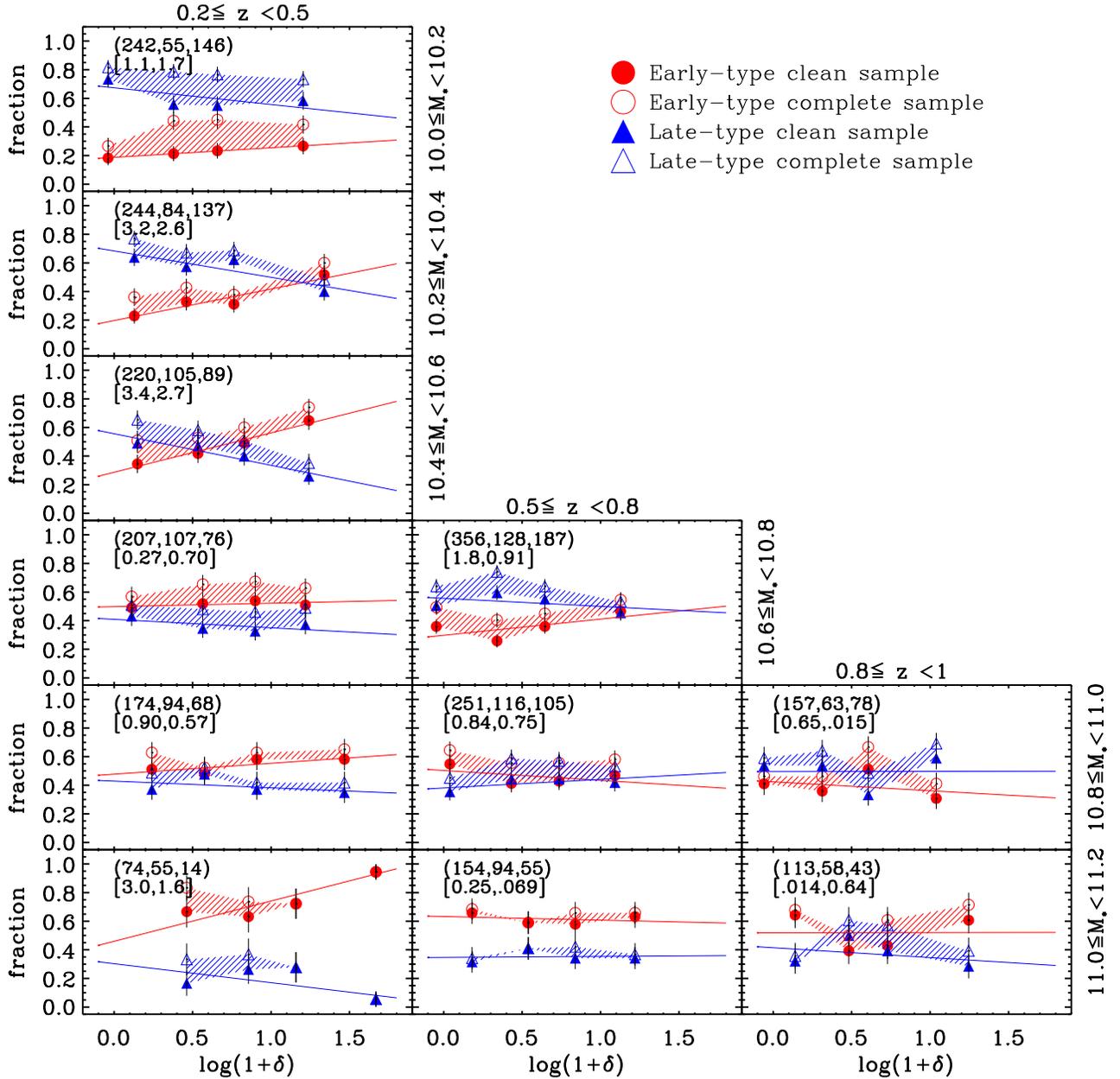}}
   \caption{Mass dependence of the morphology--density relation. The relation between the fraction of
   galaxies of different morphological types and the mass--weighted galaxy overdensity computed 
   using the $5^{th}$ nearest neighbour density estimator with volume--limited tracers is shown in
   redshift and mass bins. From the left to the right the three columns correspond to higher 
   redshifts as quoted in the label on the top. From top to bottom we consider objects of
   increasing mass as quoted in the label on the right--end side. 
   Filled red circles and filled blue triangles represent the more conservative samples of 
   early and late--type galaxies, respectively.
   Empty red circles and empty blue triangles represent the more complete, 
   but also contaminated, samples of early and late--type galaxies, respectively.
   All points are plotted as a function of the median overdensity in the respective quartile 
   (see text for more details).
   The shaded areas represent the uncertainty in the morphological classification, 
   while the vertical error
   bars are the standard fraction errors given by binomial statistics.
   Enclosed in parenthesis are the total number of objects in the mass and
   redshift bin considered, as well as the number of the more conservative sample of
   early and late--type galaxies.
   Within square brackets we give the significance of the deviation from zero
   of the slope of the linear fits relative to the clean early and late--type 
   galaxies, respectively. 
   }  
   \label{fig:Fig6}
   \end{figure*}

Firstly, we look whether the morphology--density relation is still present as
a function of redshift up to z $\sim1$ and then whether there is a possible 
dependence of the morphology--density relation on galaxy luminosity. 
While it is largely accepted that galaxies with different morphologies
and stellar populations preferentially reside in different environments, it has 
not yet been shown whether this segregation is related to the galaxy luminosity.

Using a subset of the same COSMOS imaging data used here, Guzzo et al. (2007) 
have shown that at z $\sim 0.7$ the morphology-density relation depends on 
luminosity (cf. their Figs. 11 and 12).
Here we can explore this evidence over a much broader range of redshifts and 
luminosities and in particular verify its role in creating the relation itself. 
For this pourpose we use homogeneous and complete morphology subsets of 
galaxies. 
The completeness is guaranteed by the selection of luminosity volume--limited 
samples, as explained in Section~\ref{sec:subd3} and shown in 
Figure~\ref{fig:Fig2}. 
We define luminosity volume--limited subsamples using evolving absolute 
magnitude cuts. 
We divide our main sample in three redshift bins ($0.2 \le z <
0.5, 0.5 \le z < 0.8$ and $0.8 \le z < 1$) and the statistics is 
sufficiently high to subdivide each of them in absolute magnitude bins.
The homogeneity is a consequence of the fact that galaxy morphology is measured
on the HST/ACS mosaic with same method in the whole redshift range considered.

Our findings on the luminosity dependence of the morphology--density relation 
up to z $\sim1$ are shown in Figure~\ref{fig:Fig3}.
Higher redshift bins are considered while moving to the right--end side of
the plot, while luminosity is increasing from top to bottom.
On each panel galaxies are divided into four equally populated density bins and the
median value of the density in each interval is plotted.
Results are shown for the two samples discussed in Section~\ref{sec:suba1}:
filled symbols refer to the more conservative and clean 
samples, while empty symbols are used for the complete and more contaminated ones.
The shaded regions quantify the uncertainty in the morphological classification.

Fixing the redshift, i.e. looking at Figure~\ref{fig:Fig3} vertically, we
observe a clear morphology--luminosity dependence, with a clear increase with 
luminosity of the fraction of early--type
galaxies relative to the fraction of the later population.
Spiral and irregular galaxies (grouped in the late--type population, blue
triangles) are dominant at fainter absolute magnitudes and 
in general more abundant than ellipticals.
Also we observe that the relation between the morphology of
a galaxy and the environment tends to flatten out towards fainter magnitudes 
suggesting
that brighter galaxies, at a fixed redshift, are the ones for which the
morphological segregation is stronger.

Since we use evolving absolute magnitudes to define our volume--limited 
subsamples, Figure~\ref{fig:Fig3} can be read horizontally to test how the
morphology--density relation evolves for a galaxy population with the same
luminosity.
We consider now the luminosity at which the morphology--density relation, at a
fixed redshift, flattens. We observe that for higher redshift galaxies 
the luminosity at which galaxies of a given morphological type have the 
same probability to live in an overdense or underdense region is brighter 
with respect to low redshift galaxies.
We finally point out that for galaxies with the same evolving liminosity the 
morphology--density relation is clearly present at low redshift, tends to 
flatten at intermediate redshift and almost disappears at high redshift.
All the described relations are seen for both the clean and the 
complete samples, even if we observe that the significance of the described 
relations results to be higher for the clean early--type and the corresponding 
complete late--type samples.

The use of a logarithmic scale allows the direct visualisation of 
fractional changes in the abundance of a population as a function of 
environment, irrespective of its global abundance.
For this reason, even if we do not show the equivalent of Figure~\ref{fig:Fig3}
in logarithmic scale, we summarise in Figure~\ref{fig:Fig15} the main results 
obtained by fitting the morphology--density relation with linear and 
logarithmic scales.
This figure shows the slope of the morphology--density ralation as a function
of galaxy luminosity, in three different redshift intervals ($0.2 \le z <0.5, 
0.5 \le z < 0.8$ and $0.8 \le z < 1$).
In the top row each point represents the slope of the linear fit to the 
morphology--density relation for early--type galaxies, as already shown in 
Figure~\ref{fig:Fig3}. 
The absolute value of the slope for late--type galaxies is not shown since it
does not differ, because their behaviour with the environment is complementary 
to the early--type one.
This is not the case in logarithmic scale.
We therefore represent on the central and bottom row the slope of the linear
fit to the morphology--density relation derived when the fraction of 
galaxies (late--type in the middle row, early--type in the bottom row) is 
plotted in logarithmic scale.

In the upper row of Figure~\ref{fig:Fig15} we plot the aforementioned slopes, in
the case of a linear scale, as a
function of galaxies absolute magnitude and we observe a
decrement in the slope at fainter absolute magnitudes at all redshifts.
We show this trend for early--type galaxies only, since as said above, the 
late--type population behaves in the same way. 
The only difference is in the inversion of the points
representing the clean (filled circles) and the complete (empty circles)
populations.
The second row of Figure~\ref{fig:Fig15} shows the relation with the absolute 
magnitude of
the slopes obtained from the linear fit of the morphology--density
relation in logarithmic scale for late--type galaxies.
For these galaxies the environment acts more strongly on the brightest objects 
at all redshifts.
The three panels of the third row show the same relation as in the second row,
but for early--type galaxies.
We note that in the redshift interval $0.2 \le z < 0.5$ there is an inversion
of the slope behaviour as a function of luminosity with respect to the trend
observed for the linear scale: the slope gets steeper towards faint 
luminosities.
This is also the redshift--luminosity bin where the fraction of late--type 
galaxies is the highest $(60-80\%)$. 
As a consequence, a small fractional change in the abundance of late--type 
galaxies as a function of environment, is enough to steepen the logarithmic 
slope of the early--type galaxies. 
Because of the strong uncertainty in the slope estimation, the statistical 
significance of the shown trends is not very high, but we clearly see hints for 
a trend that we plan to explore with higher statistics using the zCOSMOS 20k 
bright sample.

We finally want to stress that while we present our results using density
contrasts estimated for the smallest scale allowed by the survey design, we 
performed the same analysis using also larger scale lengths to look for a 
specific scale that might affect galaxy properties more than others.
This aspect is thoroughly examined in Cucciati et al. (2009).
For the purpose of this paper it is enough to mention that the 
morphology--density relation survived mostly unchanged when using overdensity 
estimators on larger scales. 
The only difference is seen for the highest density quartiles in the
direction of an enhanced effect when using the smallest scale,
corresponding here to the $5^{th}$ nearest neighbour.
By using this estimator we probe the
densest region of groups where the morphology--density relation is expected to
be stronger. 

   \begin{figure*}[ht!]
   \centering
   \resizebox{!}{10cm}{\includegraphics{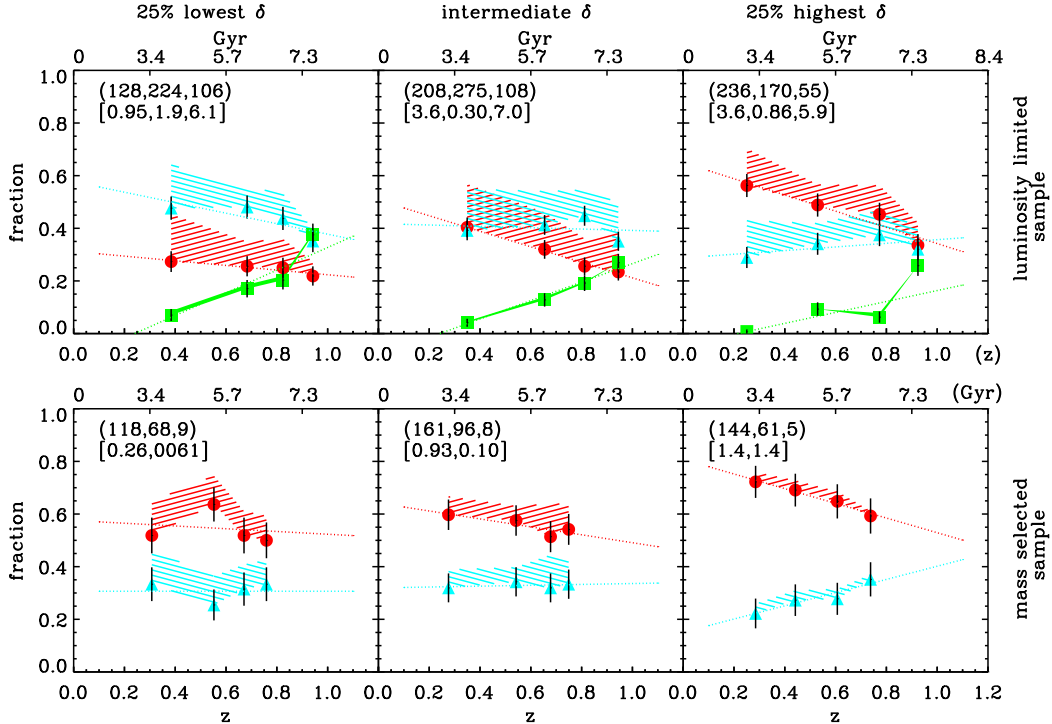}}
     \caption{Redshift evolution of the morphological fraction of galaxies as a
     function of the environment. 
     In the attempt to follow the evolution of the same galaxy population, an 
     evolving--luminosity volume
     limited sample of galaxies with $M_B < -20.5-z$ is used in the top three
     panels, while a volume--limited, stellar--mass selected sample of galaxies with
     masses $\ge  10^{10.8} M_{\odot}$ is used on the three bottom panels
     (see text for more details).
     From the left to the right we explore environments of increasing density.
     In each panel the filled red circles represent
     early--type population, the filled cyan triangles spiral galaxies and the
     filled green squares irregular objects. 
     The shaded regions represent the uncertainty in the morphological 
     classification 
     while vertical error bars are the standard fraction errors given by 
     binomial statistics.
     The linear fits of the morphological fraction of galaxy types as a 
     function of redshift for the more conservative samples are shown by the 
     dotted lines.
     Enclosed in parenthesis are the total number of early--type, spiral and
     irregular galaxies in the considered environment.
     Within square brackets we give the significance of the deviation from
     zero of the slope of the linear fits for the three galaxy 
     populations studied. 
     The evolution of irregular galaxies for the mass--selected sample has not
     been shown due to the small number of these galaxies present in these bins.    
     }
      \label{fig:Fig9}
   \end{figure*}


\subsection{Morphology--density relation: mass dependence} 
\label{sec:subr2}

   \begin{figure}[h!]
   \resizebox{\hsize}{!}{\includegraphics{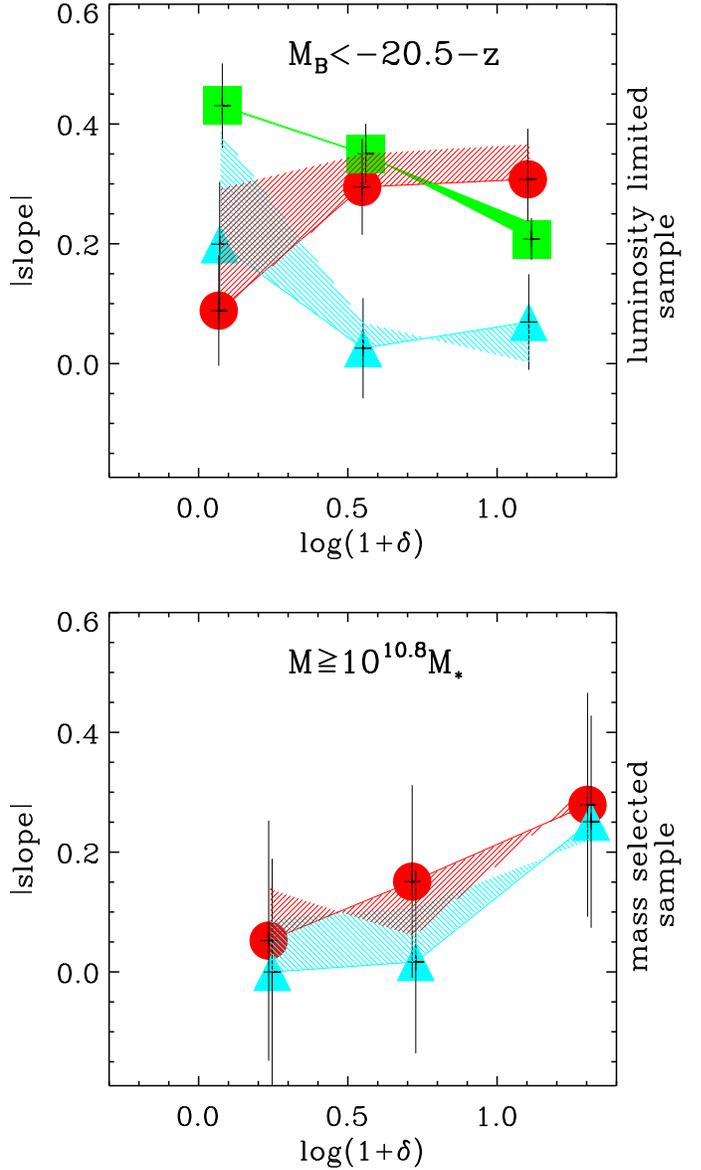}}
     \caption{The best fit slope of the fraction of early (red filled circles),
     spiral (cyan filled triangles) and irregular (green filled squares)
     population as a function of redshift, and its associated 1$\sigma$ error 
     bar, is shown as a function of the same three overdensity bins
     of Fig.~\ref{fig:Fig9}. The shaded areas quantify the uncertainty in the
     morphological classification.
     The upper and bottom panels correspond respectively to the best fit slope
     for a luminosity--selected, volume--limited sample of galaxies with 
     $M_B < -20.5-z$ and 
     a volume--limited, stellar--mass selected sample of galaxies with masses 
     $\ge  10^{10.8} M_{\odot}$.
     The trend relative to irregular galaxies in the mass--selected sample is 
     not shown due to the small number of irregular objects with such a
     high mass.         
     }
      \label{fig:Fig14}
   \end{figure}

Galaxies show a significant B--band rest--frame luminosity evolution between 
redshift zero and one, which has been shown to be dependent on the spectral 
(Zucca et al. 2007) and morphological (Zucca et al. 2009) types.
In addition, due to bursts of star formation,  the B--band rest--frame 
luminosity can strongly change during galaxy life.

In literature it is often claimed that galaxy stellar mass varies much 
less during galaxies life and can be considered as a more stable quantity to 
characterise a galaxy than its B--band rest--frame luminosity.
Bolzanella et al. (2009) show that the galaxy stellar mass function depends on
the environment. The strong variation of mass--to--light with stellar population
age, and thus with the star formation history of a galaxy, together with the
galaxy mass segregation produces a range of masses in a luminosity limited
sample.   
Stellar mass should therefore represent a more physically motivated parameter on
the basis of which to select galaxies, to better understand the evolution of
galaxies populations.

We investigate in this Section whether the strong relations 
that galaxy morphologies have with the environment in luminosity--selected 
volume--limited samples are still present when using stellar--mass volume 
limited samples.
This approach allows us to tackle the issue of whether the aforementioned effects
are simply due to a biased view induced by the luminosity selection or a 
more fundamental relation among galaxy stellar mass, environment and morphology
does exists.
In other words, we will try to answer to the following question: 
are the physical processes which determine the mass more important than
the environment in shaping galaxy morphology? 

For this analysis we use volume--limited stellar--mass complete galaxy subsamples 
as explained in section~\ref{sec:subd3} and shown in Figure~\ref{fig:Fig5}.
The main sample is divided in three redshift bins (the same used for the study
of the luminosity dependence) that we now subdivide in mass bins 
sufficiently narrow not to mix effects related to the mass, while still having
a significant statistics.

Figure~\ref{fig:Fig6} shows our results on the mass dependence of the
morphology--density relation.
As in Figure~\ref{fig:Fig3}, the redshift increases from the left to the right 
panels, while stellar masses increase from top to bottom. 
In each panel densities have been divided into equally populated quartiles and 
the median value of the density in each interval is plotted.
Filled  and empty symbols in the various panels correspond respectively to the
clean and complete samples defined in Section~\ref{sec:suba1}.
The significance of the deviation from zero of the slope of the linear fit of 
the morphology--density relation is indicated inside the square brackets in 
each panel of Figure~\ref{fig:Fig6} for the clean early and late--type samples 
respectively.
We observe that for all the redshift and stellar mass bins considered the
significance of the slope of the fit does not depend on the sample used to
compute it: a relation which is significant at more than $2-3 \sigma$ with the 
complete sample stays significant with the clean one. 

We find that for galaxy stellar masses $\gtrsim 10^{10.8}$ M$_{\odot}$
the probability of a specific morphological type to reside in a low or high 
density environment is approximately the same. 
Expressed differently, the morphology--density
relation substantially weakens at intermediate/high masses. 
The only possible exception is for the highest mass bin at low redshift (lower
left panel) where a possible dependence of galaxy morphology on the environment
appears to be present. 
Nonetheless we note that also this morphological trend with respect to the
environment would be flat if we do not consider the last quartile corresponding
to the densest environment.
We assume that we are dealing with a residual relation and we explore whether 
the reason is the contamination of the early--type population by
late--type galaxies and/or the incompleteness of late--type galaxies.
In this panel the morphology--density relation is significant at $3 \sigma$ for
the clean sample of early--type galaxies (or complete sample of late--type) and
is mostly consistent with being flat for the complete sample of early--type (or
clean sample of late--type) objects.
We conclude that subsequent studies with higher statistics are needed to confirm
the behaviour of the morphology--density relation for very massive galaxies at 
low redshift.

At stellar masses smaller than $\sim 10^{10.6}$ M$_{\odot}$, which are probed by
our mass--complete sample only in the first redshift bin, the fraction of 
early--type galaxies increases towards denser regions, while the late--type 
galaxies are more abundant in a sparse environment than in denser regions.
The low significance of the morphology--density relation in the upper--left 
panel, corresponding to the lowest stellar masses and redshifts considered in 
this analysis, can be interpreted in at least two different ways. 
It could be due to the presence, in this mass and redshift interval, of a small 
degree of incompleteness
which acts in the direction of a flattening of the relation but it could also be 
a real effect indicating that at low mass the role of the environment becomes
again secondary with respect to the mass.
If confirmed, this would support the hypothesis that there is a restricted range 
of stellar masses for which galaxy morphology has an environmental dependence.
Since at stellar masses lower than $\sim 10^{10.0}$ M$_{\odot}$ we quickly 
become incomplete we conservatively conclude that the morphology--density 
relation does not vanish for low mass galaxies and that the inversion of 
behaviour we witness, with respect to higher stellar mass objects, could 
be seen as an indication of the existence of
a critical mass which separates galaxies into two distinct families, in the
framework of what already shown at low redshift by Kauffmann et al. (2003). 

Finally reading the plot horizontally, 
approximatively equivalent to following  the evolution in time of the 
morphological properties of galaxies with a given mass
(de Lucia et al. 2006; Pozzetti et al. 2007), 
we qualitatively observe 
the evolution with redshift of the galaxy morphological types, for galaxies with
the same mass.
We witness a progressive increase with cosmic time of the fraction of 
early--type galaxies in the same mass bin. 
This can be seen as the consequence of the transformation from 
late to early--type galaxies.
The redshift evolution for very massive galaxies, the only ones for which we are
complete up to z=1, is then quantitatively described in the bottom panels of 
Figure~\ref{fig:Fig9}, that we fully describe in the following Section.

The main scenario resulting from our analysis is already well--drawn from our
combined study of the morphology--density relation for luminosity and
stellar--mass selected, volume--limited samples.
The morphology--density relation is present out to z $\sim1$ at fixed
luminosity, but its behaviour gets flatter when we look at it at fixed masses
expecially above $10^{10.6}$ M$_{\odot}$.
This is a clear indication that at least above $10^{10.6}$ M$_{\odot}$ the
role of  stellar mass is dominant with respect to the environment in 
determining galaxy morphology.
We now want to discuss how the observed differences in the morphological 
segregation as a function of luminosity and stellar mass can be explained.
We discuss two possible explanations which can be seen as extreme cases:
\begin{itemize}
\item it is a fact that for galaxies less massive than $10^{10.6}$ M$_{\odot}$ 
the morphology--density relation exists.
If we suppose that the galaxy stellar mass function is the same in all
environments then, for the lower M/L objects, a given luminosity bin could reach 
down to these lower mass bins. 
As a consequence, because of the mixing of masses, we see a residual 
morphology--density relation at fixed luminosity;
\item  Bolzonella et al. (2009) show that the galaxy stellar mass function
varies with the environment with more low mass galaxies being present in low
density environment.
Even if the morphology--density relation were flat for all masses,
which is not what we observe but can be taken as an extreme case,
more low mass galaxies (most of them spirals) in low density
regions than in high density ones would enter any given luminosity bin.
The final effect would be that of producing a morphology--density relation at 
fixed luminosity.
The existence of a mass segregation, such that more massive galaxies with older 
stellar population, which mainly have early--type morphologies, preferentially 
reside in denser regions, together with the strong relation of morphology, colour
and M/L to mass may be the main driver in producing the observed 
morphology--density relation.
\end{itemize}

Even if both scenarios might partly contribute to the observed effect, we 
consider the second hypothesis as the more important, since the variation with
the environment in the global galaxy stellar mass function (Bolzonella et al. 
2009), coupled with the strong correlation between
morphology and mass, appears to be large enough to produce the observed 
effects.
In addition the first hypothesis presented is a "mixing effect", mixing
together equal numbers of high and low mass objects in all environments and
averaging out their individual morphology--density relations.
While we claim that there is an effect below $10^{10.6}$ M$_{\odot}$, we are not
overwhelmed by the strength of the morphology--density relation, at fixed mass,
in this regime. Averaging it out, as mentioned above, it is just going to make 
it weaker still. On the other hand our second hypothesis does not even need a
variation at all. The variation of the galaxy stellar mass function seems large
enough over the range of ten or so needed in mass to produce the observed 
effect, when coupled with the strong dependence on morphology with mass.


\subsection{Evolution of the morphology--density relation} 
\label{sec:subr3}

The understanding of the role of the environment in the determination of the
morphological mix of galaxies observed at various redshifts is fundamental 
for our comprehension of galaxy formation and evolution.
We analyse in this section the evolution of the morphological mix of galaxies as
a function of the environment for luminosity and stellar--mass selected, volume 
limited samples.
In Section~\ref{sec:subr2} we have shown that the morphology--density relation
observed in luminosity--selected, volume--limited samples can be due to 
the biased view imposed by the B--band luminosity selection.
The observed trends are mainly due to the fact that luminosity--selected samples
include the low mass, bright blue galaxies, while they miss the equally low 
mass, red counterparts.
Being aware that more physically motivated results come from the use of 
stellar--mass selected, volume--limited samples, we still consider in the
following analysis both luminosity and stellar--mass selected, volume 
limited samples, mainly to allow the comparison to previous studies.

When looking at the observed fraction of bright galaxies of each morphological 
type as a function of redshift (i.e. Zucca et al., 2009), we observe 
that the bright late--type population becomes increasingly dominant at higher 
redshifts, while the fraction of bright early--type galaxies decreases.
To follow the same galaxy population at different redshifts we therefore correct
for evolution using a luminosity--selected, volume--limited sample with a cut--off 
which evolves with luminosity. 
In particular, to track the evolution of the morphology--density relation over 
the redshift range $0.2 \le z < 1$ we select objects with M$_{B_{AB}}<-20.5-$z 
as these galaxies are visible over the entire redshift range.

The redshift evolution of the morphological mix in different environments
is shown in Figure~\ref{fig:Fig9} for evolving--luminosity (upper panels) and 
stellar--mass selected (lower panels), volume--limited samples. 
Figure~\ref{fig:Fig14} instead shows the growth rate for early-type, spiral and
irregular galaxies as a function of
the environment. 

The upper panels of Figure~\ref{fig:Fig9} 
show that in the redshift range $0.2 \le z < 1$ the evolution of the 
relative fraction of early--type, spiral and irregular galaxies depends on the 
environment.
Little evolution is seen in the early--type population in low density 
environments, while we see a monotonic rise with
cosmic time in the highest density regions. 
We can confirm with our data the
behaviour found by Smith et al. (2005) at intermediate densities, where the
evolution seems to occur quite recently with little evidence for any change at
earlier epochs.
When focusing on spiral galaxies, the lack of evolution at
intermediate and high density regions is striking. 
Some degree of evolution in the 
fraction of spiral galaxies is instead visible in low density environments.
To counterweight the evolution observed in the early--type population we expect a
strongly evolving trend for irregulars, that is in fact what is found.
Figure~\ref{fig:Fig9} finally confirms that the relative fraction of 
morphological types changes with redshift and environment. 

Galaxy luminosities are very sensitive to bursts of star
formation, which are frequent at the high redshifts considered.
Since stellar masses are expected to change less rapidly with redshift, we opt 
for the use of mass--selected samples.
In doing that, we simply cut the luminosity--selected, volume--limited samples 
at $10^{10.8}$ M$_{\odot}$.
The results are shown in the second row of Figure~\ref{fig:Fig9}.
The trends we observe for this sample are extremely different from the ones 
seen for the luminosity--selected, volume--limited sample. 
Irregular galaxies are not plotted here due to their very small number at such
high stellar masses. 
In the lowest and intermediate density environments we witness a trend
consistent with lack of evolution for both early--type and spiral galaxies.
Some indications of evolution, although at low statistical significance, less 
than the $2 \sigma$ level, is present in the densest regions.
As already shown in Section~\ref{sec:subd3} and Figure~\ref{fig:Fig4} the main 
difference between the evolving--luminosity and the stellar--mass selected 
samples is in the presence in the first sample of low mass, star forming 
galaxies which are not included in the mass--selected sample.
This suggests that these objects as the main responsibles of the evolution 
seen in the luminosity volume--limited samples.
			    
We define the growth rate of the fraction of a specific morphological type in a
given environment as the best fit slope of the morphological mix evolutionary
trends shown in Figure~\ref{fig:Fig9}. 
The timing of environmental evolution of the galaxies morphological mix can 
be investigated just looking at Figure~\ref{fig:Fig14}, where the growth
rate for each morphological type considered in this study is shown as a 
function of the environment.
In Figure~\ref{fig:Fig14} we plot the absolute value of the best fit slope, which
can be taken as a measure of the efficiency of the morphological evolution.
For luminosity volume--limited galaxy samples, strong evolution is seen in low 
density environments for spirals and irregulars, in contrast to the small 
evolution of early--type galaxies. 
The trend is clearly inverted in denser environments.
In the case of mass--selected samples, almost no evolution is seen
independently of the environment and/or of the morphological type, with the
possible exception only in the highest density environment.
 
Our results are based on the assumption that the luminosity evolution is the 
same for all galaxies independently of their morphology, which might not be the 
case.
As control check we therefore define evolving absolute magnitude cuts, 
different for each of the 3 morphological classes considered, and we follow the
redshift evolution of the relative population in different environments.
To parametrise the evolution we use the values of M$^{\ast}$ obtained in Zucca 
et al. (2009) in a B rest--frame band with $h=0.7$ and with a fixed 
$\alpha$ defined in the redshift range $0.3 < z < 0.8$. 
The evolution of M$^{\ast}$ is then fitted with a simple linear function for
each morphological type.
Despite the fact that the value of M$^{\ast}$ varies as a function of the
galaxy morphological type
we observe that the final general trend does not change, confirming that the 
result is robust and does not strongly depend on the specific value assumed for 
the luminosity evolution.

%

\section{Morphological mix of red and blue galaxies} 
\label{sec:specT}
   \begin{figure}[h!]
    \resizebox{\hsize}{!}{\includegraphics{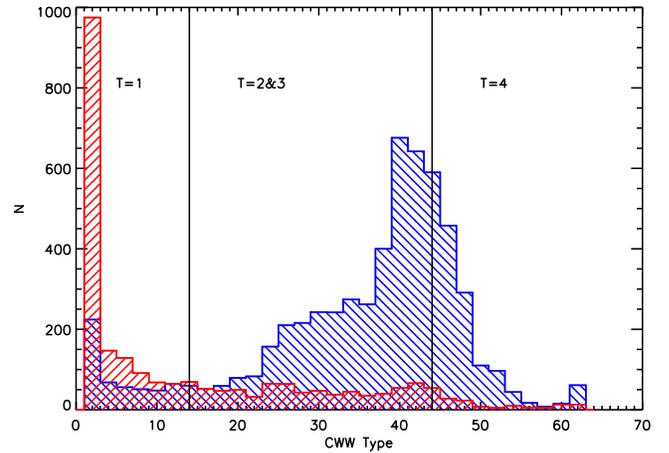}}
      \caption{Distribution of the extended CWW templates as a function of the morphological type.
      The early--type distribution is represented by the red
      ($45^{\circ}$--inclined) shaded histogram,
      while the late--type population is shown by the blue ($-45^{\circ}$--inclined) shaded
      histogram.
      Vertical lines separate the spectrophotometric types: T=1,2,3,4 corresponds to E/S0, 
      early spiral, late spiral and irregular/starburst templates, respectively.
              }
         \label{fig:Fig7}
   \end{figure}

   \begin{figure*}[ht!]
   \centering
    \resizebox{!}{19cm}{\includegraphics{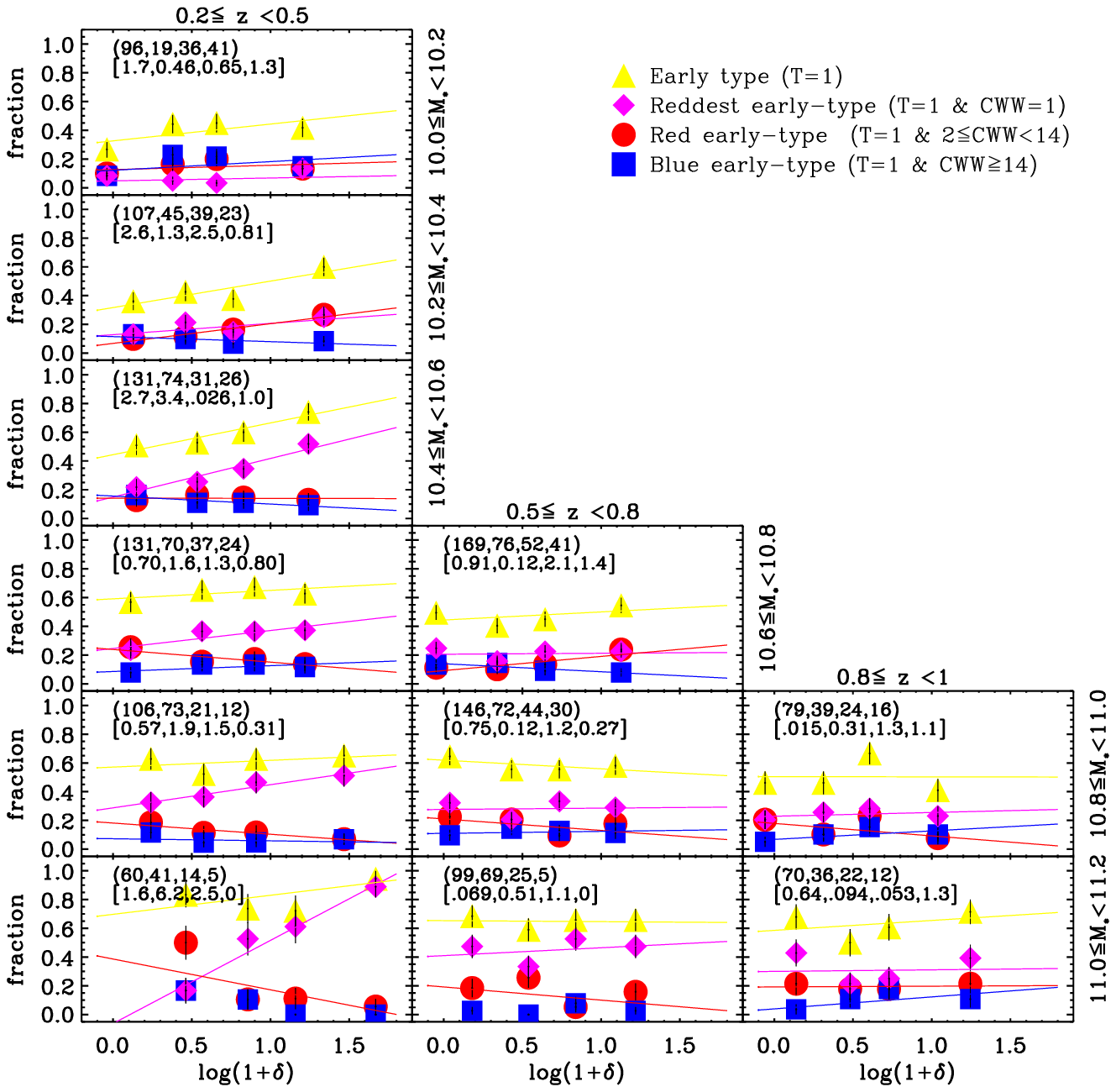}}
     \caption{Contribution of the various spectrophotometric types to the early--type population
     in different environments. 
     The percentage of the whole morphologically classified early--type population, as well as the 
     percentage of the spectral type subclasses,
     is plotted as a function of the mass--weighted galaxy overdensity computed using the $5^{th}$
     nearest neighbour density estimator with volume--limited tracers. The redshift increases
     from the left to the right (as indicated on the top), while objects are progressively 
     more massive from top to bottom (as indicated on the right--end side). 
     The entire early--type population is represented with filled yellow triangles.
     This population is then divided into 3 subclasses according to the galaxy spectral type: 
     filled magenta diamonds correspond to very red spectral type ($CWW=1$); filled red circles 
     represent E/S0 fitted by the templates 2--13; blue early--types ($CWW \geq 14$) are represented 
     with blue filled squares.  
     Enclosed in parenthesis are the total number of galaxies in the stellar mass and
     redshift bin considered, as well as the number of very red, red and blue early--type 
     objects.
     Within square brackets we give the significance of the deviation from
     zero of the slope of the linear fit. 
     	    }     
      \label{fig:Fig12}
   \end{figure*}

   \begin{figure*}[ht!]
   \centering
   \resizebox{!}{19cm}{\includegraphics{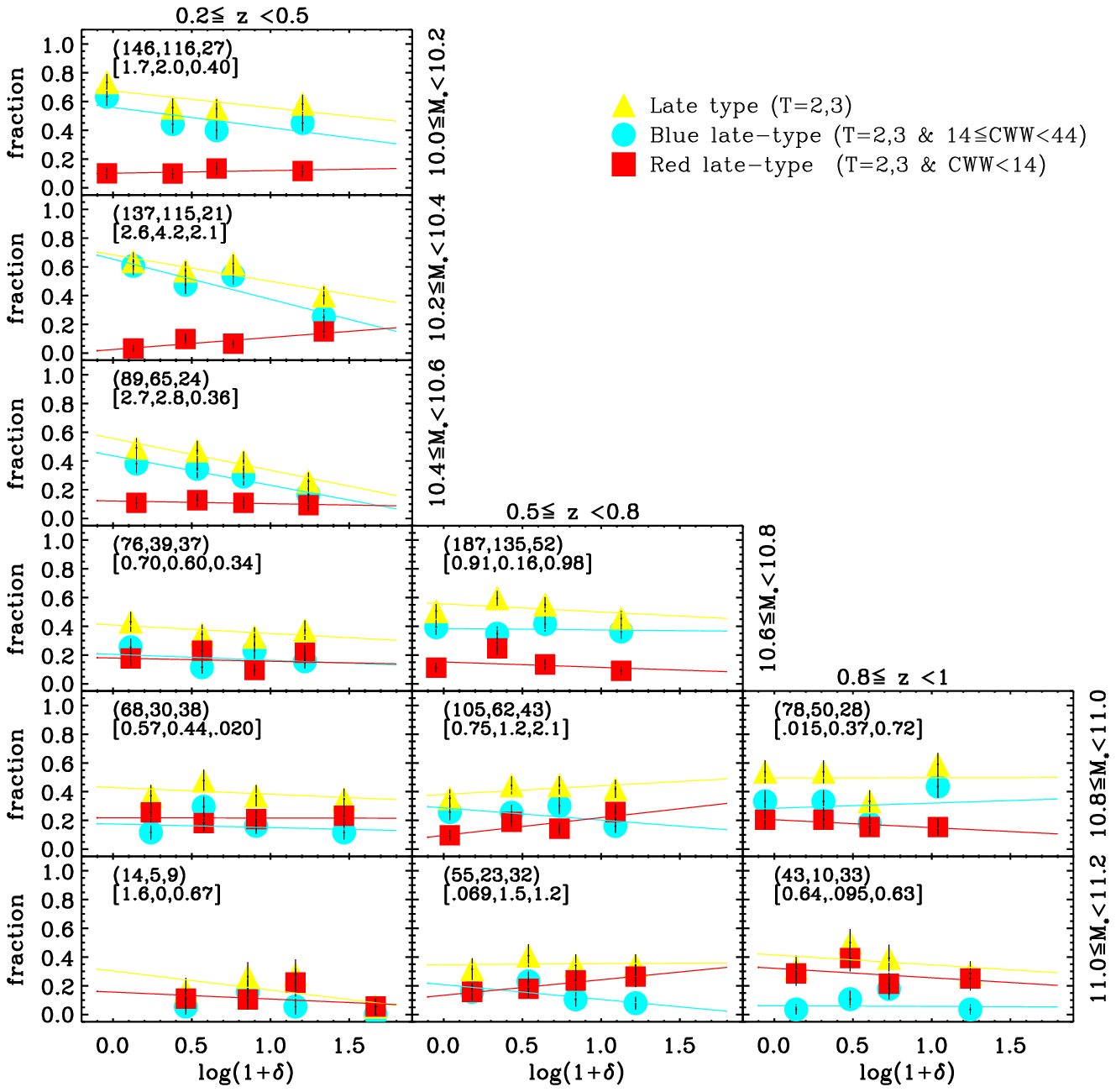}}
     \caption{Contribution of the various spectrophotometric types to the late--type population
     in different environments.  
     The percentage of the whole morphologically classified late--type population, as well as the 
     percentage of the spectral type subclasses,
     is plotted as a function of the mass--weighted galaxy overdensity computed using the $5^{th}$
     nearest neighbour density estimator with volume--limited tracers. The redshift increases
     from the left to the right (as indicated on the top) while 
     objects are progressively more massive from top to bottom (as indicated on the right). 
     The entire late--type population is represented with filled yellow triangles.
     This population is then divided into 2 subclasses according to the galaxy spectral type: 
     filled cyan circles represent early and late spirals fitted by the templates 14--43; 
     red late--types ($CWW \leq 13$) are represented with red filled squares.  
     Irregular and starburst spectral types ($CWW > 43$) are not shown, since they are not
     present at the masses considered. 
     Enclosed in parenthesis are the total number of objects in the mass and
     redshift bin considered, as well as the number of blue and red late--type 
     objects.
     Within square brackets we give the significance of the deviation from
     zero of the slope of the linear fit. 
     	    }
      \label{fig:Fig13}
   \end{figure*}

In this section we further explore the contribution of different stellar 
populations to the morphology--density relation in order to disentangle 
dynamical and star formation history contributions 
and better study the relative importance of the environment and of the physical
mechanisms which determine galaxy stellar masses to the galaxy
segregation.

The galaxy morphology could be viewed as the end result of its
dynamical history.
Frequent high speed galaxy encounters (galaxy harassment)
drive the morphological transformation of galaxies in clusters, leading to 
dramatic evolution of disk galaxies (Lake et al. 1998). Ram pressure
stripping and tidal stripping also play a key role in transforming galaxies in 
clusters centres.
In our survey we do not sample such rich clusters, which are rare such that
galaxies in this environments represent only 1\% of the population at z $\sim0$.  
Galaxy mergers are believed to be the primary mechanism that creates 
ellipticals, if the merger involves two spiral galaxies that have approximately 
the same mass (i.e. major merger) and happens in a way that 
drives away much of the dust and gas through a variety of feedback mechanisms.
 
By definition, galaxies spectrophotometric types depend on galaxy colours,
which themselves depend on the recent star formation. 
To assign our galaxies to spectrophotometric classes, we have fitted their 
observed multi--band photometry with the spectral energy distribution of 
sixty--two CWW (Coleman et al. 1980) extended
templates plus two starburst templates (Kinney et al. 1996).
Galaxies are then divided into four spectrophotometric types, according to their
spectral energy distribution from UV to near--IR, corresponding to 
E/S0 (type 1), early spiral (type 2), late spiral (type 3) and irregular
(type 4) templates. The two starburst templates are included in type 4.
As shown in Zucca et al. (2006) going from type 1 to type 4 objects the composite
spectra have a bluer continuum with increasingly stronger emission lines.

The existence of high--resolution HST/ACS imaging and multi--wavelength coverage
of the COSMOS field allows to combine and compare morphological and 
spectrophotometric information.
We name early--type galaxies with a spectrophotometric type 2, 3 or 4 $(CWW \ge
14)$ "blue",
while the ones with spectrophotometric type 1 are named "very red" when fitted
by the most extreme type 1 template $(CWW=1)$ and "red" when fitted by the other 
type 1 templates $(2 \le CWW < 14)$. 
The same logic is used to classify late--type objects in a "red" population,
when fitted by type 1 templates $(1 \le CWW < 14)$, in a "blue" population,
when the spiral templates of type 2 and 3 $(14 \le CWW < 44)$ are the best fit 
of the galaxy SED and in a "very blue" population when fitted by the more
extreme starburst templates $(CWW \ge 44)$  corresponding to type 4.
 
Figure~\ref{fig:Fig7} shows the distribution of early and late--type galaxies 
as a
function of the spectrophotometric templates. The $45^{\circ}$ inclined red 
shaded histogram represents the morphological early--type galaxies, while the
$-45^{\circ}$ inclined blue shaded one represents the late--type galaxies in
our sample. 
We observe a tail in the histogram of early--type galaxies extending bluewards, 
up to the irregular and starburst spectrophotometric templates.
A visual check of these objects reveal that $\sim37\%$ of them are face--on 
late--type galaxies with morphological parameters typical of an 
early--type population (for more details, see Cassata et al., in preparation).
The existence of a dominant bulge and the consequent small contribution of the 
disk component in the estimation of the structural parameters used in the
automatic galaxy classification procedure is one of the reasons that we bring 
forward to explain why those objects are automatically classified as
early--type.
In order to explore more in detail this population, stacked spectra have been
produced following Mignoli et al. (2009). 
The composite spectra of "blue" early--types present prominent emission lines
(Zucca et al. 2009): 
low mass "blue" early--type galaxies have stacked spectra, typical of star
forming galaxies, with blue continuum and emission lines; the composite spectra
of more massive "blue" early--types galaxies show emission lines on top of
a red spectra with absorption lines.
For massive "blue" early--type galaxies, the presence of a central AGN could
explain the high concentration we measured for these galaxies on HST/ACS images. 
Evidence for blue colours or star formation signatures in the
spectra of early--type galaxies has been reported by various studies
in the past (Schade et al. 1999; Im et al. 2001; Menanteau et al. 2001, 2004; 
Cross et al. 2004; Cassata et al. 2005; Treu et al. 2005a; Cassata et al. 2007).   
The latter paper in particular studied in detail the properties of the blue 
early--type population in a sub--set of the COSMOS ACS data centered around 
z $\sim0.7$, showing in particular that their fraction is independent 
of local density.

We note that $\sim 13\%$ of late--type galaxies populate the tail of the
histogram extending redwards.
Furthermore, we note that the peak corresponding to the reddest template 
$(CWW=1)$ contains a considerable fraction $(\sim 16\%)$ of extremely red
late--type galaxies. 
We decided to visually inspect all the morphologically late--type objects with 
a SED better fitted by a red template to be reassured that our population of 
red late--type galaxies is real and not simply due to an error in the 
morphological classification.
We find in this class of objects a well characterised 
population of edge--on
spiral galaxies, often dominated by a strong dust line and therefore highly
reddened. This is in agreement with what already found in the DEEP survey by
Weiner at al. (2005), in the  GEMS survey by Bell et al. (2004b) and in the
COSMOS field by Cassata et al. (2007).
The latter paper in particular is based on a subset of the same ACS 
observations used here and discusses in detail how this population of 
highly--inclined dust--reddened spirals represents a $\sim 30\%$  
contamination in a colour--selected sample of supposedly passive red galaxies.

Once this careful check performed,  we create the "clean" and 
"complete" samples, used in our analysis and introduced in
Section~\ref{sec:suba1}. 
They are defined in this way: the "clean" sample is the one
for which blue ellipticals are removed from the morphologically classified 
early--type population, the spiral class staying unchanged; instead in the 
"complete" sample the blue ellipticals are added to the morphologically
identified spiral galaxies.

\section{colour dependence of the morphology--density relation} 
\label{sec:subc4}

If we allow for the existence of an environmental dependence of the global
stellar mass function (Bolzonella et al. 2009) and of strong mass--morphology,
mass--density and mass--M/L relations, the morphology--density relation at
constant luminosity could be interpreted as due to selection effects.
We therefore consider only stellar--mass selected, volume--limited samples 
to investigate the role of the star formation history on the morphology--density
relation. 
We start our investigation from the morphology--density relation shown in 
Figure~\ref{fig:Fig6} and we combine morphological and spectrophotometric 
information.

In Figure~\ref{fig:Fig12} we show the contribution of different
spectrophotometric types to the early--type population, going from sparse to
high density regions, as a function of galaxy stellar mass.
The blue early--type galaxies show a behaviour with respect to the density 
either consistent with being flat or decreasing with density, typical of 
a late--type population, 
even if always at less than $2\sigma$ significance. 
This is expected since, as mentioned above, this population is largely composed
by face--on, bulge--dominated spirals.
We now compare the relative contribution of very red and red early--type
galaxies to the slope of the morphology--density relation in different
stellar mass and redshift bins.
We observe that at z $<0.5$ and for all the stellar mass bins considered, the 
morphology--density relation for the very red galaxies (magenta diamonds) is 
present at few $\sigma$ level, while the contribution of the other 
spectrophotometric types to the global trend (yellow triangles) is pretty 
negligible and consistent with being flat.

In Figure~\ref{fig:Fig13} we show the contribution of different
spectrophotometric types to the morphology--density
relation of the late--type population as a function of galaxy stellar mass.
The contribution of starbursts and irregulars becomes negligible at the
considered stellar masses and it is therefore not shown.
The panels of Figure~\ref{fig:Fig13} show how the population of red 
late--type galaxies, that is in part produced by the spurious 
effect of internal reddening,  is substantially independent of local density 
(apart from local fluctuations due to small number statistics as in the 
lowest--redshift, lowest--mass bin).  
This is expected, as it only depends on the galaxy inclination with respect to 
the line--of--sight (see also Cassata et al. (2007) for a direct measurement of 
the dependence of colour on galaxy inclination).  
Thus, the overall late--type population is intrinsically described by blue 
templates only.  
These fully drive the observed lack of dependence on local density for massive 
galaxies and the limited dependence at small masses and low redshift.
The comparison in each single panel of Figure~\ref{fig:Fig13} of the 
significance of the slope of the linear fit of the morphology--density relation 
for different spectrophotometric types shows that either no morphology--density
relation is present for red late--type objects (red squares) 
or that this population has a preference for high density regions, like 
early--type galaxies.
The spiral spectrophotometric types (cyan dots) are 
responsible for the observed late--type galaxy segregation at masses smaller than
$10^{10.6}$ M$_{\odot}$. 

The interpretation of Figure~\ref{fig:Fig12} and ~\ref{fig:Fig13}, where we 
investigate how the star formation history affects the morphology--density 
relation as a function of galaxy stellar mass, could be done in terms of the 
aging of the stellar population. 
We already claimed from Figure~\ref{fig:Fig6} that there is no morphological
segregation, in the galaxy population as a whole, at least for masses larger 
than $10^{10.6}$ M$_{\odot}$ at any redshift.
In Figure~\ref{fig:Fig12} we observe that at z $<0.5$ and at fixed mass, within the
unsegregated early--type population, there is a systematic colour/age segregation:
redder and therefore older galaxies (magenta diamonds) show a stronger 
dependence on the environment than the younger population (red dots).
This behaviour seems to soften at higher redshifts, where the dependence on the
environment of the spectrophotometric types (of which the early--type population
is the global representation) is consistent with being flat.
An age dependence on the environment is also visible in Figure~\ref{fig:Fig13},
where blue and red spirals often present an opposite trend with environment.
In the Sloan Digital Sky Survey (SDSS) (Blanton et al. 2005; Quintero et al.
2006) and more recently in the GalaxyZoo project (Bamford et al. 2009) it is
found that galaxy colour is more strongly correlated with environment than 
morphology and that colour and morphology bimodalities are largely independent
functions of environment.
Our results show the same but, for the first time, at high redshift.
This result, combined with the lack of morphological segregation for high mass
galaxies, is an indication that morphology is a galaxy scale property while star
formation is more affected by large scale structure.

%

\section{Discussion}
\label{sec:discussion}


A strong consequence of the cold dark matter paradigm in cosmology is the
hierarchical scenario of galaxy formation according to which haloes are 
assembled from smaller units through mergers. 
A number of studies confirm that galaxies, located in these dark matter haloes, 
are complex systems and their evolution definitely depends on the
interplay of various factors such as star formation history, chemical
enrichment, feedback and dynamical effects, including mergers. 
All these parameters have their importance, at different times in the life of
a galaxy, in shaping the great variety of objects we observe.

In addition to this work, a set of related studies, also based on zCOSMOS
spectroscopy and COSMOS photometry, approach this topic using a variety of
statistical tools (Bolzonella et al. 2009; Pozzetti et al. 2009; Zucca et al.
2009) and analysis (Cucciati et al. 2009; Kova\v{c} et al. 2009; 
Iovino et al. 2009)
to understand how star forming history, stellar mass, environment and merger 
history (de Ravel et al. 2009b; Kampczyk et al. 2009; Tresse et al. 2009) 
differently act to determine the properties of individual galaxies.

\subsection{Comparison to previous studies} 
\label{sec:subc1}

Significant progress has been made in the understanding of the
environment--morphology connection since Dressler et al. (1980) first claimed
that a relation exists between the morphology of a galaxy and the environment in
which it resides. 
A quantitative comparison of our results with previous 
analyses is not straightforward mainly because of the different morphological
classification schemes and density estimators adopted in the various surveys.
Compared to many previous studies, we have the advantage to use a surface 
density estimator (number of galaxies in cylinder divided by the circular area, 
but with the real distances)
which properly corrects for all the survey observational biases 
(Kova\v{c} et al. 2009).
We also fully exploit the possibility of accurate morphological classification 
over an unprecedented volume provided by the COSMOS HST/ACS imaging 
(Scoville et al. 2007; Koekemoer et al. 2007), building upon previous 
morphological analyses of the same data set (Capak et al. 2007; 
Cassata et al. 2007; Guzzo et al. 2007). 

Cooper et al. (2005) claim that environmental investigations
require good precision spectral measurements to reconstruct the galaxy
distribution down to small scales.
The early environmental analyses of the COSMOS data presented in Capak et al. 
(2007), Cassata et al. (2007) and Guzzo et al. (2007) show that spectroscopic 
redshifts
are needed to recover density/structures in low density environments, which is
where a large fraction of galaxies live. However, in high density regions the
environment can be properly recovered using high quality photometric redshifts. 
Large and deep redshift surveys are therefore the best probes of the density
field delineated by galaxies at all redshifts since they definitely increase the
dynamic range of density which can be recovered.
In this respect, the zCOSMOS survey was designed to
reconstruct the 3D density field out to z $\sim1$, from 100 kpc to 100 Mpc scale,
with much better resolution in the radial direction than it has been possible
with photometric redshifts or weak lensing.

The main observational obstacle in probing the structure of high redshift
galaxies is the seeing--limited and limited spatial resolution of ground--based
images which can introduce significant classification biases, ie. galaxy
structures such as spiral arms and tidal tails get progressively smoothed out
with increasing redshift.
Among the large number of quantitative classifiers that have been developed
or extended over the years we opt in this study for an innovative 
non--parametric approach (see Section~\ref{sec:suba1} for more details) 
which allows for a uniform classification up to z $\sim1$.

Galaxy spectrophotometric types are often used instead of morphological types
since they can be measured directly from multi--wavelength photometry.
In addition, the results interpretation is strictly related to colour studies and
benefits of interesting literature (Cooper et al. 2006; Cucciati et al. 2006).
However, we clearly observed that morphological and spectral types do not
fully overlap and complementary information are obtained when using both of
approaches, as in this work.
colour studies are more related to the galaxy stellar population and 
its star formation history, while morphological studies tell us more about the 
effects of dynamics and environment on galaxy evolution.

Recent relevant works (Smith et al. 2005; Capak et al. 2007; Guzzo et al. 2007; 
Van der Wel et al. 2007) look at the effects of the environment on morphological 
evolution up to z $\sim1.2$.
In many of these studies the environment is computed following the method first
introduced by Dressler et al. (1980) and often adopted later on 
(Dressler et al. 1997; Postman et al. 2005). 
The projected number density is calculated by counting the 10 nearest neighbors 
and dividing by the rectangular area enclosed enclosed by them.
Contamination from the projection of field galaxies at lower and higher
redshifts is corrected in Smith et al. (2005) using Postman et al.'s (1998)
I--band number counts and in Capak et al. (2007) using photometric redshifts
bins for the nearest neighbor counting.
Van der Wel et al. (2007) also used local projected surface densities, but
measuring the distance to the $n$th nearest neighbor more massive than 
$4x10^{10}$M$_{\odot}$ to be more consistent with their choice to work with mass
selected samples.
The morphological classification instead is quite inhomogeneous in its 
definition through the various studies and sometimes within the same study.
An exception is the work by Capak et al. (2007), who introduce the use of 
the Gini coefficient measured in a Petrosian aperture as a stand alone 
parameter able to clearly separate early and late type galaxies with
$0.3<$z$<1.2$ using the ACS F814W filter.
Smith et al. (2005) visually classified 1257 galaxies in high density clusters
and low density fields starting from HST/ACS observations at z $\sim1$. 
To discern
evolutionary trends they compare with Dressler (1980) at z $\simeq0$ and with
Dressler et al. (1997) and  Treu et al. (2003) at z $\simeq 0.5$. 
van der Wel et al. (2007) automatically classified their sample of more than
2000 galaxies using the S\'ersic parameter $n$ and a measure of the residual 
($B$) from the model fit (Blakeslee et al. 2006).  
They claim that their morphological 
classification of galaxies has been quantified in an internally consistent 
manner both in the local universe and at high redshift and that the $n-B$
classification is the condition to achieve this result.
The reason is that they take into account the PSF smearing effect and this is 
essential when a sample with very different photometric properties is used to 
study redshift dependent trends.

In their study on the COSMOS field, Guzzo et al. (2007) found that at z=0.73 the
morphology--density relation was already in place with a globally lower fraction 
of early--type galaxies ($\sim 65\%$) than at z=0 for comparable local
densities, while Capak et al. (2007) conclude that galaxies
are transformed from late to early type more rapidly in dense than in sparse
regions and that no evolution at all is found at z $>0.4$ at densities below 100
galaxies Mpc$^{-2}$. 
Their findings are qualitatively in agreement with Smith et al. (2005) who
show that the early type fraction is $\sim3$ times steeper locally
than at z=1 and that most of the evolution responsible of the observed local
trend occurred at z $<0.5$.
Van der Wel et al. (2007) opt for an innovative approach and get as main
result that the fraction of early type galaxies more massive than
$4\times10^{10}$M$_{\odot}$ in the field and group environment has remained
constant since z $\sim0.8$ and that therefore the galaxies evolution in mass, 
morphology and density should happen without changes in the morphology--density 
relation.

Where overlapping, our results are in agreement with previous works on the
evolution of the morphology--density relation and possible small differences 
can be explained by the involved approach.
Nonetheless it should be stressed that the strength of our results is
additionally supported by the homogeneity of the sample used for the analysis.
Using luminosity and stellar mass volume limited samples, splitted in five
redshift intervals up to z$\sim1$, we looked at information on morphological
evolution rates for galaxies above some limiting luminosity or stellar mass.
We additionally looked at the difference between colour and morphology--density
relation. 
Previous studies have the merit for already exploring the behaviour 
of certain galaxy properties in well defined environments, often clusters,
and in specific luminosity, mass and redshift ranges
But it is the first time that a comprehensive analysis is done using a 
complete, well-checked and uniform set of spectroscopic and photometric data.
For this unique characteristic of our data, the scenario they describe allow 
to tie together previous findings in a simple general view.


\subsection{Role of the environment on galaxy formation and evolution} 
\label{sec:subc2}

How do galaxies form and what determine their evolution?
Is the destiny of a galaxy established by birth or are external factors
important? 
Which is the role of the environment in the complex scenario we can put 
together using the large number of observational and theoretical analyses
carried out in the last decades?
Is the morphology--density relation an intrinsic galaxy property, "nature"
hypothesis, or is due to the environment, "nurture" hypothesis?

From an observational point of view,
the downsizing in galaxy formation (Cowie et al. 1996; Gavazzi et al. 1996) 
predicts that more
massive galaxies cease to form stars at early epochs, while less massive systems
stay active longer.
This apparent anti--hierarchical scenario is also supported by other studies in 
the local universe (e.g. Kauffmann et al. 2003) and at high redshift 
(e.g. Glazebrook et al. 2004).
In addition, Juneau et al. (2005) show that the more massive galaxies start to 
form stars earlier than intermediate and low mass objects.
More massive galaxies also acquire galactic structures like bars earlier, 
showing that dynamical maturity of disks also follows cosmic downsizing 
(Sheth et al. 2008).
Finally, Einasto et al. (2005) claim that clusters in high density regions
evolve more rapidly than in low density environments, supporting the theoretical
expectation of an acceleration of structure formation in denser environments.

From a theoretical perspective,
Frenk et al. (1985) performed the first numerical simulation of a flat CDM 
universe and
claimed that the assumed hierarchical clustering model is consistent with the 
observed dynamics of galaxy clustering only if galaxy formation is biased 
towards high density regions.
Therefore, they first proposed that the morphology--density
relation is a natural consequence of a hierarchical scenario of galaxy 
formation.
As shown, among others, by Frisch et al. (1995), the first objects to form in
simulations are rich clusters in superclusters.
More recently, high--resolution simulations and semi--analytic models of galaxy
formation (de Lucia et al. 2006) confirm the trends of deep high redshift
surveys.

Because of the properties of our data, the following considerations on the 
mechanisms involved in the formation and evolution of galaxies are focused on 
the history of massive galaxies from z $\sim1$ to present.
We start from a scenario, supported by numerical simulations and ultra deep
observations, in which the early universe is mainly populated by spirals and 
irregulars. 
Various studies claimed that massive early type galaxies are already in place 
at z=1.
De Ravel et al. (2008), using VVDS data, show that major mergers account for 
about $20\%$ of the mass assembled in present day galaxies with masses $>
10^{10}$ solar masses, the rest being due to minor mergers and passive 
evolution.
Early type galaxies should therefore form either by environmental effects such 
as merging and interaction, or simply by stopping star formation after losing 
or exhausting their gas.
Moreover some early type galaxies may have formed since the beginning 
as ellipticals and not from the transformation of spirals and irregulars.

Intuitively, the most massive galaxies at any epoch should be the final product of
the evolutionary chain: either they assemble very fast by mergers or they form
very early.

We already discussed that the evolutionary scenario we can deduce from the study
of the morphology--density relation in luminosity--selected samples may be
largely a consequence of the biased view imposed by our B--band luminosity 
selection.
Nonetheless, we explored the results of this approach for comparison to previous 
works which adopted the same luminosity selection.
The stellar--mass selection is instead a more physical approach on the basis of
which to identify the key players in galaxy evolution.

From our analysis we observe that when considering a luminosity volume--limited 
sample, the fraction of early type galaxies changes as a function 
of the environment up to z $\sim1$: in high density regions we observe a 
larger fraction of early type objects compared to low density environments 
(see Figures~\ref{fig:Fig3} and~\ref{fig:Fig9}).
We interpret the observed morphological segregation as the combined result
of the existence of an environmental dependence of the global stellar mass
function (Bolzonella et al. 2009) and of the well--known mass--morphology and
mass--M/L relations. 

The volume--limited, stellar--mass selected sample used in this study allows 
considerations on the evolution of galaxies more massive than
$10^{10.6}$M$_{\odot}$ and on the role of less massive objects.
The existence of a critical mass above which we do not observe any morphological
dependence with respect to the environment (see Figure~\ref{fig:Fig6}) suggests
that, at least for what concerns massive galaxies, the intrinsic 
mechanisms of galaxy formation might be more fundamental than the environmental 
processes.
In addition, the comparison of the flat slope of the morphology density relation 
seen in a stellar--mass volume--limited sample with the trends existing 
when using a luminosity volume--limited 
sample brings to the conclusion that blue galaxies less massive than
$10^{10.6}$M$_{\odot}$, for which our B--band luminosity selection misses the red
counterparts, are likely to be the main contributors to the evolution
observed in the luminosity volume--limited sample. 

Which scenarios could be advocated to speculate on the causes of the surviving
of a morphology--density relation at masses lower than $10^{10.6}$M$_{\odot}$ and
for the existence of the differences in the galaxy stellar mass function for low
and high density environments?
We consider the following alternatives:
\begin{itemize}
\item starting from a uniform distribution of galaxies, there has been a
more effective transformation in high than in low density environments.
The "nurture" scenario is preferred.
\item since the beginning there are more ellipticals in denser regions. 
If "nature" acts alone, the fraction of early type galaxies should stay mostly
constant with time. 
We should therefore combine this hypothesis with the one that
the highest density regions are the first to collapse and have consequently more 
time to increase their early type population by merging and environmental 
effects.
\end{itemize}
The environment also affects the growth morphological populations: the 
early--type fraction increases faster in dense than in sparse regions, the 
opposite happening for spirals.
Within the limits of our statistics, irregulars seem not to feel the
environment, they strongly evolve in redshift regardless of the environment
(see Figure~\ref{fig:Fig14}).
These results point towards an environmental dependence of the different
processes governing galaxy formation, which can preferentially act on high or 
low density regions.

Another ingredient which contributes to the general picture and that we can
extrapolate from our data is that morphology and star formation appear to be
affected by different processes.
We show that at high galaxy stellar masses morphology is not driven by the 
environment, while galaxy colours are still affected by it (see
Figure~\ref{fig:Fig12}).
Is the growth in the early type fraction mainly driven by interactions, while 
the reduction in star formation is caused by gas stripping?

We conclude that both "nature" and "nurture" play, at different epoch in the 
history of a galaxy, an important role in shaping the large variety of objects 
we observe in the local universe as final result of the evolutionary chain.

%

\section{Summary \& conclusions}
\label{sec:final}

We use the so--called 10k sample, constituted by the first 10644 observed
galaxies with spectroscopic redshifts of the "bright" zCOSMOS survey, to study 
the evolution of the morphology--density relation up to z $\sim 1$ for
luminosity and stellar mass selected, volume--limited galaxy samples.
We use the density estimates of Kova\v{c} et al. (2009) who introduce a new 
algorithm, implemented in the ZADE code (Kova\v{c} et al., 2009), to derive the 
3D density field.
This new algorithm makes advantage of the high sampling rate of our 
spectroscopic sample as well as of the high accuracy
of the photometric redshifts measured in the COSMOS field.
The high resolution and depth of the HST/ACS imaging available in the field is
essential to perform an objective, automatic and unbiased morphological
classification. It is the first time that the morphological mix can be split 
into early type, spiral and irregular galaxies up to z $\sim 1$ for such a 
large sample. 
Finally, the unique multi--$\lambda$ coverage of the COSMOS field is essential
for an accurate estimate of galaxy stellar masses and spectrophotometric types.
The COSMOS and zCOSMOS surveys are therefore the first surveys to enable the 
study up to z $\sim1$ of the impact of the environment on galaxy formation and 
evolution, trying to disentangle nature and nurture effects, using a complete 
and homogeneous sample with high quality control in spectral measurements.

In agreement with previous studies (Postman et al. 2005; Smith et al. 2005; 
Capak
et al. 2007) we see that the morphology--density relation, in luminosity
selected volume--limited samples, was already in place at z=1, but tends to be 
flatter than at lower redshift. 
This is also in agreement with the observed trends in the colour--density
relation (Cooper et al. 2006; Cucciati et al. 2006). 
In addition, we observe a luminosity dependence of the morphology--density
relation: the morphological segregation is stronger for the brightest galaxies 
at each redshift. 
Nonetheless, at least at the lowest redshifts considered for our analysis, the 
environment acts more effectively on low luminosity early--type galaxies, 
affecting more their star formation and morphology in dense regions. 
This behaviour is reversed for late--type objects suggesting a different role 
of the environment on the morphological mix at the same luminosity.
Looking at the dependence with respect to the local overdensity of the 
spectrophotometric types relative to the early--type population we conclude 
that the brightest and reddest galaxies at each redshift are responsible of 
the global trend observed.

zCOSMOS enables us to investigate whether the strong effects we are witnessing 
in luminosity 
selected volume--limited samples are simply due to a biased view induced by 
the B--band luminosity selection or a more fundamental relation 
among galaxy stellar masses, environment and morphology does also exist.
In fact, the morphology--density relation observed in the luminosity--selected 
sample can be considered as a "selection effect" arising from the existence of
an environmental dependence of the galaxy stellar mass function (Bolzonella et 
al. 2009) and the well--known mass--morphology and mass--M/L relations.
We additionally observe, when using luminosity--selected, volume--limited samples, 
a strong environmental dependence in the evolution of early--type and spiral 
galaxies in the explored redshift range: the behaviour of the early--type
population with redshift is consistent with no evolution in low density regions,
while showing a monotonic increase with cosmic time at intermediate and high 
overdensities; the evolution with redshift of spiral galaxies as a function of
the environment has the opposite trend with respect to early--type galaxies.  
In contrast irregulars seem to strongly evolve regardless of the local 
environment.

When considering stellar--mass selected, volume--limited samples the 
aforementioned  trends are much less significant, supporting the hypothesis 
that the 
observed evolution is due to low mass, bright, star forming galaxies for which
the B--band luminosity selection misses the equally low mass red counterparts.

We stress the importance of using a stellar--mass selected, volume sample in 
order to obtain an information less biased by the galaxy star formation history 
and more physical.
Is there a critical mass which separates galaxies into two families having a
different behaviour with respect to the environment? Is the galaxy stellar mass
or the environment the primary driver of the morphological differentiation? 
Is the mass a more fundamental parameter than the environment in which a galaxy
lives? 
Our data show that at masses lower than $10^{10.6} M_{\odot}$ the
morphology--density relation is present at $\sim 3 \sigma$ level. 
At higher stellar masses we witness a change in this behaviour, with galaxies of 
a specific morphological type not showing any environment dependence.
Since galaxy stellar mass is the result of several processes, including
conditions at birth, this result seems to indicate that the physical processes 
that determine the morphology of massive galaxies are rather independent of 
the environment or that at least at z $<1$ massive galaxies evolve 
independently of the environment.

Interestingly we still observe a colour segregation for a specific
morphological type in mass regimes where no morphology--density relation can be
claimed.
We then conclude that there is a colour dependence on the environment beyond that
on morphology up to z $\sim 0.5$, with some hints at higher redshifts, most 
likely related to environment dependent star formation histories.

Even if the zCOSMSOS 10k sample allowed for a great improvement and refinement
in our understanding of the role of the environment in shaping the Universe we
observed, there are questions which are still without a precise answer mainly 
due to statistical limitation and to the limit in redshift imposed by the
relatively bright limiting magnitude. 
The 20k zCOSMOS "bright" sample, that we are finishing assembling, as well as
higher redshift data from the zCOSMOS deep project should overtake these
shortcomings and allow to extend our analysis to an epoch where important 
changes occur and to firmly establish the redshift at which the morphology
density relation emerges.

\begin{acknowledgements}
      LT acknowledge support from CNES. 
      We thank CNES and PNC for support to the COSMOS project. 
      This work has been partially supported by INAF grant 
      PRIN-INAF 2007 
      and by the grant ASI/COFIS/WP3110I/026/07/0.

\end{acknowledgements}

\end{document}